\documentclass[useAMS,usenatbib]{mnras}
\usepackage[utf8]{inputenc}
\usepackage[usenames]{color}
\usepackage{ulem}
\usepackage{graphicx}   
\usepackage[T1]{fontenc}
\usepackage{ae,aecompl}

\usepackage{newtxtext,newtxmath}

\usepackage{graphicx}	
\usepackage{amsmath}	
\usepackage{amssymb}	

\def\apj{ApJ}
\def\apjl{ApJL}
\def\aap{A\& A}

\def\mnras{MNRAS}

\def\apjs{ApJS}

\def\mnras{{MNRAS}}

\title[Properties of HI structures]{The physical and the geometrical properties of simulated cold HI structures
}

\author[A. Gazol \& M. A. Villagran]{Adriana Gazol$^{1}$ and Marco A. Villagran$^{2}$
\thanks{E-mail: a.gazol@irya.unam.mx}
\\
\\
$^{1}$Instituto de Radioastronom\'ia y Astrof\'isica, UNAM, campus Morelia. PO Box 3-72. 58090. Morelia, Michoac\'an, M\'exico
\\
$^{2}$Instituto de Astronom\'ia y F\'isica del Espacio, UBA-CONICET, Ciudad Universitaria. C1428ZAA. Bs. As., Argentina
}

\date{Accepted XXX. Received YYY; in original form ZZZ}

\pubyear{2019}

\begin{document}
\label{firstpage}
\pagerange{\pageref{firstpage}--\pageref{lastpage}}
\maketitle

\begin{abstract}
{The objective of this paper is to help shedding some light on the nature and the properties of the cold structures formed via thermal instability in the  magnetized atomic interstellar medium. To this end, we searched for clumps formed in forced (magneto)hydrodynamic simulations with an initial magnetic field ranging from 0 to 8.3$~\mu$G. We statistically analyzed, through the use of Kernel Density Estimations, the physical and the morphological properties of a sample containing $\sim 1500$ clumps, as well as the relative alignments between the main direction of clumps and the internal velocity and magnetic field. The density ($n\sim 50-200$~cm$^{-3}$), the thermal pressure ($P_{th}/k\sim 4.9\times 10^3-10^4$~K cm$^{-3}$),  the mean magnetic field ($\sim 3-11$~$\mu$G ), and the sonic Mach number of the selected clumps have values comparable to those reported in observations. We find, however, that the cloud sample can not be described by a single regime concerning their pressure balance and their Alf\'enic Mach number. We measured the morphological properties of clumps mainly through the asphericity and the prolatness, which appear to be more sensitive than the aspect ratios. From this analysis we find that the presence of magnetic field, even if it is weak, does qualitatively affect the morphology of the clumps by increasing the probability of having highly aspherical and highly plolate clumps by a factor of two, that is by producing more filamentary clumps. Finally, we find that the angle between the main direction of the clumps and the local magnetic field lies between $\sim\pi/4-\pi/2$ and shifts to more perpendicular alignments as the intensity of this field increases, while the relative direction between the local density structure and the local magnetic field transits from parallel to perpendicular.
}
\end{abstract}



\section{Introduction}

It is well known that magnetic field is a crucial agent in determining the structure, dynamics, and evolution of diffuse ISM \citep[e.g.][]{Heiles2012,HennebelleInut2019FrASS...6....5H}. The study of the intrinsic anisotropic nature of its effects together with the multi-phase constitution and turbulent character of the material, have stimulated abundant and varied approaches involving observations, theory and numerical simulations. In recent years, the study of the interplay between the magnetic field and the structure of interstellar clouds has been encouraged by the arrival of observations which strongly suggest a close relationship between them.

In fact, observations have revealed an apparent pervasive presence of filament networks in the cold interstellar medium (ISM), both in molecular clouds \citep[e.g.][]{Williamspp2000prpl.conf...97W, Andrrepp2014prpl.conf...27A, Alina2019MNRAS.485.2825A} as well as in cold HI gas   \citep[e.g.][]{McClure2006ApJ...652.1339M, Clark2014ApJ...789...82C, PlanckHI2016A&A...586A.135P, kalberlaetal2016ApJ...821..117K,kalberlahaud2018, Verschuur2018}. 
In the later case, a relationship between the filaments and the magnetic field has been pointed out in a variety of contexts. In particular, observational studies have reported  a seeming  preference for a local alignment between the two dimensional density structures and the magnetic field projected on the plane of the sky \citep{Clark2014ApJ...789...82C, Zaroubi15, kalberlakerp16, PlanckHI2016A&A...586A.135P, kalberlalerphaudhav2017,jelicetal2018}.

The relative orientation between cold HI structures and the magnetic field has been studied by distinct groups though different methods: the Rolling Hough Transform has been used for high spatial and spectral resolution HI data combined with starlight polarization data \citep{Clark2014ApJ...789...82C} and with dust polarization data \citep{Clarketal2015}, as well as for numerical simulations of multi-phase gas \citep{Inoue_2016}. The Histogram of Relative Orientations (HRO), first proposed by \cite{Soler2013}, has been employed to analyse polarization and intensity maps from the dust emission at 353 GHz detected by Planck \citep{PlanckHI2016A&A...586A.135P}, which presumably  corresponds to diffuse gas that can be identified with the Cold Neutral Medium (CNM, $T\sim 50-200$~K, $n\sim 10-100$~cm$^-3$). 
In \cite{Marco18} we studied the HRO in the cold neutral gas resulting from multi-phase magnetized models with a scale of 100\,pc, finding that there is a preferred alignment between the CNM structures and the local magnetic field for initial magnetic field intensities between 0.4 and 8$\mu G$, but this preference weakens as this intensity increases.

Even though the cold HI gas is easily detected, mainly through the 21 cm line but not only, the actual measure of its physical characteristics as density, temperature, thermal pressure, velocity dispersion, or magnetic field properties, requires both, high quality observations and complicated data analysis \citep[e.g.][]{Heiles2003, HeilesTroland2004,HeilesTroland2005,Jenkins2011, Murray2017, Goldsmith2018, ClarkHensley2019, Syedetal20}. As a consequence, observational constraints to the physical properties of the CNM are limited and concern mainly the solar neighborhood.  In particular, this is the case for the magnetic field intensity because its measurement through the 21 cm line Zeeman splitting represents a mayor observational challenge \citep[see][and references therein]{HeilesCrutcher05}. Thus, the current knowledge about the magnetic field intensities in the CNM  comes mainly  from the Millenium Arecibo 21 Centimeter Absorption-Line Survey \citep{HeilesTroland2004, HeilesTroland2005}. 

On the other hand, despite the extensive work concerning the properties
of individual CNM structures on various scales, only few of them have been  morphologically characterized \citep[see e.g.][and references therein]{Heiles19}. And despite the existence of recent  and promising techniques to trace the magnetic field structure in the diffuse interstellar medium, \citep[][]{GonCasLaz2017, Tritsis2018}, which have already been successfully applied to the HI gas \citep[][]{Yuenetal2017, Huetal2019, Tritsis2019},  the theoretical predictions concerning the nature and the extension of the connection between the CNM clump morphology and the magnetic field still scarce.  

In this context, numerical models offer the possibility to assess the relationship between the properties of CNM-like structures and its magnetic field. Thus, the main objective of the present work is to provide a detailed statistical analysis of the possible connection between the general physical properties, the morphological properties, and the geometrical properties of magnetized CNM-like structures. As a consequence, we are interested in study density structures resulting from the development of the condensation mode of thermal instability \citep{Field65} in a magnetized medium with turbulent motions. To this, end we analyse clumps formed in forced simulations with thermodynamic properties similar to those of the local HI gas and five different initial magnetic fields. In order to characterize the clump morphology, and in addition to the traditional aspect ratio, we explore a more sensitive shape descriptor frequently used in physical chemistry known as asphericity, which has the further advantage of being easily related to the two dimensional aspect ratio. Regarding the geometry of clumps, we use directional statistics to study the relative orientation of the dense structures with respect to the magnetic field and to the velocity field as well as the relative orientation between these fields. 

In past years, the morphological properties of density structures resulting from the development of thermal instability in a magnetized gas have been numerically studied by \cite{Hennebelle2013}, and \cite{wareing16,Wareing2019}.  However, in addition to the different analysis tools, our approach is complementary to those presented by these authors. When compared  with \cite{Hennebelle2013} our work explores the effect of having different ambient magnetic field strengths; while in contrast to \cite{wareing16,Wareing2019} our aim is to analyse dense structures with properties similar to those observed in the CNM rather than filaments formed during the gravitational contraction of molecular cloud precursors.
 
 The paper is organized as follows. In the next section we introduce the analysis tools that are used through the paper,  while the numerical models are described in Section \ref{sec:sim}. The results on the three dimensional clump morphology and geometry are then reported  and discussed in Sections \ref{sec:results} and \ref{sec:disc}. In Appendix \ref{sec:app} we explore the morphological and geometrical  properties of  two dimensional clumps selected from projections.  Finally, our conclusions are presented in Section \ref{sec:conc}.

\section{Analysis Tools}\label{sec:tools}
The simulations that we use are intended to model the interstellar atomic gas as a continuous, but in practice 
the data are discrete sets of values. For this reason we describe the analysis tools in terms of discrete quantities. 
\subsection{Useful Tensors}
For a discrete set of particles with the same mass, the gyration (or shape) tensor $\mathbf{G}$ is defined as
\begin{equation}
G_{ij}= \displaystyle\sum_k x^k_i x^k_j,
\end{equation}
where $\mathbf{x^k}$ is the vector position with respect to the centroid of the particle distribution. The eigenvalues $g_1$, $g_2$, $g_3$  of $\mathbf{G}$ represent the squared semi-axes length of an ellipsoid containing the particle distribution, and are related with the gyration radius  $r_g$ by $r_g=\sqrt{g_1+g_2+g_3}$. The corresponding eigenvectors are the principal-axes of the mass distribution.

When particles have different masses $\mu^k$, the mass distribution is characterized by the tensor $\mathbf{M}$, defined as
\begin{equation}
    M_{ij}= \displaystyle\sum_k  \mu^k  x^k_i x^k_j,
    \label{eq:defM}
\end{equation}
where $\mathbf{x^k}$ is the vector position with respect to the center of mass. 
The tensor $\mathbf{M}$ differs from the inertia tensor $\mathbf{I}$, which is defined as 
\begin{equation}
    I_{ij}= \displaystyle\sum_k  \mu^k (\delta_{ij}|\mathbf{x^k}|^2 - x^k_i x^k_j). 
\end{equation}
These two tensors are however related by
\begin{equation}
 \mathbf{I}=\mathrm{Tr}\hspace{1pt}(\mathbf{M})\mathbf{I_0}-\mathbf{M},
\end{equation}
where $\mathbf{I_0}$ is the identity matrix, implying that they have the same eigenvectors (principal axes) and that if $m$ is an eigenvalue of $\mathbf{M}$, then $\mathrm{Tr}\hspace{1pt}(\mathbf{M})-m$ is eigenvalue of  $\mathbf{I}$.

The shape tensor, introduced by \cite{2011ApJS..197...30Z}, is defined as
\begin{equation}
    S_{ij}= \frac{\displaystyle\sum_k  \mu^k  x^k_i x^k_j}{\displaystyle\sum_k  \mu^k},
\end{equation}
and constitutes a normalized version of  $\mathbf{M}$.

With a consistent choice of the weighting and the normalization, the eigenvalues of $\mathbf{M}$ (and $\mathbf{S}$) are also proportional to the squared length of an ellipsoid's semi-axes containing the mass distribution.

The projection of $\mathbf{M}$ on the plane of the sky can be written as:
\begin{equation}
     P_{ij}= \displaystyle\sum_k  \Sigma  X^k_i X^k_j,
\label{eq:defP}    
\end{equation}
where $\Sigma$ is the surface density, and $X_1=x$ and $X_2=y$ are coordinates in the plane of the sky. Due
to the two-dimensional nature of $\mathbf{P}$, its eigenvalues can be simply expressed as
\begin{equation}
     p_{\pm}=P_{11}+P_{22}\pm\sqrt{P_{11}^2+4P_{12}^2-2P_{11}P_{22}+P_{22}^2}
\label{eq:eigenvP}    
\end{equation}
\subsection{Shape descriptors}\label{sec:shapedesc}
If $m_3\geq m_2\geq m_1$ are the eigenvalues of $\mathbf{M}$ and $\mathbf{S}$, then the shape of a mass distribution is traditionally measured by the ratios $\sqrt{m_2/m_3}$ and  $\sqrt{m_1/m_3}$. 

There exist, however, other shape descriptors related to the length of the principal axes.
One of them is the asphericity, introduced by \cite{Rudnick1986}, which is commonly used in physical chemistry  for the study of polymers and that measures the deviation from spherical symmetry.
 For three dimensional distributions, the asphericity is defined as:
\begin{equation}
    A_3= \frac{1}{6}{\displaystyle\sum_k \frac{(m_k-\bar{m})^2}{\bar{m}^2}},
    \label{eq:aspher_3d}
\end{equation}
where $\bar{m}=\frac{1}{3}{r_g^2}$. For spherically symmetric objects ($m_1=m_2=m_3$) $A_3=0$, while for a completely elongated object ($m_1=m_2=0$) $A_3=1$. The quantity $A_3$ is a scale invariant. In fact, from equation (\ref{eq:aspher_3d}), it can be noticed that it is a combination of two more common invariants of the matrix $\mathbf{M}$ : the trace ($r_g^2$) and the sum of minors, $\hat{m}=m_1m_2+m_1m_3+m_2m_3$,  $A=1-3\hat{m}/r_g^2$.

The asphericity measures how much the shape defined by the principal axes differs from a perfect sphere, but does not give information about how disk-like (oblate) or rod-like (prolate) is the corresponding  ellipsoid. In order to measure the sense of the deviation from the sphere, another shape descriptor, known as prolatness, can be used in addition to the asphericity.
Two definitions can be found in the literature, both of them expressed in terms of invariant quantities of the gyration tensor. In some cases \citep[e.g.][]{Blavatska2010,Aronovitz1986} the prolatness is defined as
\begin{equation}
    S_A={\displaystyle\frac{\prod_{k}(m_k-\bar{m})}{\bar{m}^3}}.
    \label{eq:prolatS}
\end{equation}
From the above definition $S_A$ can take values between $-1/4$ for perfectly oblate objects and 2 for perfectly prolate ones. An alternative but equivalent definition of the prolatness can be found in \cite{Ostermeir2010,DabrowskiTumanski2019}. 

The asphericity defined in equation (\ref{eq:aspher_3d}), also known as anisotropy, is a particular case, for a three-dimensional mass distribution, of the more general quantity
\begin{equation}
    A_d= \frac{1}{d(d-1)}{\displaystyle\sum_k \frac{(Q_k-{\bar{Q}})^2}{\bar{Q}^2}},
    \label{eq:aspher_dd}
\end{equation}
where $\mathbf{Q}$ is the gyration tensor in $d$ dimensions, $Q_k$ is the $k$-the eigenvalue of $\mathbf{Q}$,  and $\bar{Q}=\frac{1}{d}\mathrm{Tr}(\mathbf{Q})$ \citep[see e.g.][]{Diehl_1989}. From the previous expression,
\begin{equation}
    A_2= \frac{(\lambda_1-\lambda_2)^2}{(\lambda_1+\lambda_2)^2},
    \label{eq:aspher_2d}
\end{equation}
for the eigenvalues $\lambda_1$ and $\lambda_2$. If $\lambda_1$ and $\lambda_2$ are identified with $p_{-}$ and $p_{+}$,  then $A_2$ can be written as
\begin{equation}
    A_2= \frac{(r^2-1)^2}{(r^2+1)^2},
    \label{eq:aspher_2dr}
\end{equation}
where $r=\sqrt{p_{-}/p_{+}}$ is a common tool used for the characterization of  the apparent (projected) shape of a mass distribution \citep[see e.g.][]{Gammie2003ApJ...592..203G}. 
%
\subsection{Kernel Density Estimations}\label{sec:kdes}

In order to make better distinctions among the distributions resulting for the different models   
we use, instead of histograms, the probability density resulting from Kernel Density Estimations (KDEs). These estimators use a kernel function $K$ to smooth out the contribution of each data point over a local neighbourhood of that data point such that, for $n_p$ data points, the  estimated probability density at $x=x_0$  is given by
\begin{equation}
    f(x_0)=\frac{1}{n_p h}\displaystyle\sum_i K \left(\frac{x_i-x_0}{h}\right),
    \label{eq:kde}
\end{equation}
where $h$ is the bandwidth, which controls the extent of the region over which $x_0$ contributes to $f(x)$ \citep[see e.g.][]{silverman86}.
It is well known that, in general, the kernel choice has less effects on KDE results than the bandwidth choice. We use a Gaussian kernel and different values of $h$, which are determined by cross-validation for the specific data of each distribution. For this we use the cross validation algorithm \textsc{GridSearchCV} in scikit-learn \citep{scikit-learn}. Note that in section \ref{sec:results} a single value of $h$ is used in each plot. 

\subsection{Directional Statistics}\label{sec:dirstat}

In order to compute mean angles (e.g. between the longest principal axis and the local magnetic field, or between the local magnetic field and the local velocity) we use directional statistics\footnote{For more details on directional statistics we refer to \cite{Mardia99}}. Within this framework, a distribution of angles $\theta$, is considered a distribution of directed data over a circle. The mean angle is then the direction of the centre of the distribution. The mean length $\overline{R}$ of the centre of the distribution vector, which measures the concentration of the $\theta$ values, is calculated as:
\begin{equation}
  \overline{R} = (\overline{C}^2 + \overline{S}^2)^{1/2},
\end{equation}
where
\begin{align}
\overline{C} = \dfrac{1}{n}  \sum_{j=1}^{n} \cos \theta_j, \ \ \ \ \
 \overline{S} = \dfrac{1}{n}  \sum_{j=1}^{n} \sin \theta_j,
 \end{align} 
with $\theta_j$ being, in our case, the angle at each point of a clump containing $n$ voxels. For widely dispersed values of $\theta$, $\overline{R} \ll 1 $, while
for the case in which the directions $\theta_j$ are highly concentrated around a preferred direction $\overline{R} \sim 1$, meaning that the mean angle $\overline{\theta}$ is significant, and it has the form:
\begin{equation}
  \overline{\theta} =
  \begin{cases}
    \tan^{-1} (\overline{S}/\overline{C}) & \text{if } \overline{C} \geq 0,\\
    \tan^{-1} (\overline{S}/\overline{C}) + \pi & \text{if } \overline{C} < 0.
  \end{cases}
      \label{eq:theta}
\end{equation}


\section{The simulations}\label{sec:sim}
In this work we analyse data from five simulations previously presented in \cite{Marco18}, specifically runs labeled as B00S, B01S, B05S, B10S, and B20S. These models are intended to reproduce the thermal conditions of HI gas in the solar neighborhood for the purely hydrodynamic case (B00S) as well as in the magneto-hydrodynamic one (B01S, B05S, B10S, B20S). Each model represents a periodic cubic box with 100pc by side, initially at rest and with uniform density ($n_0=2$~cm$^{-3}$) and temperature ($T_0=1500$~K), in the thermally unstable regime according with a cooling function based on the results of \cite{Wolfire2003}. At the initial density and temperature, the Jeans length is 578~pc, implying that our models are highly sub-Jeans. The gas is permeated by a magnetic field initially uniform and parallel to the x-direction, which initial intensity $B_0$ varies for  each model. The values that we use for $B_0$ are 0, 0.4, 2.1, 4.2 , and 8.1~$\mu$G for B00S, B01S, B05S, B10S, and B20S, respectively. In the models, turbulent motions are induced through a Fourier space forcing with a constant kinetic energy injection rate and at a fixed wavenumber corresponding to a physical scale of $50$\,pc. The cooling function is a fit to the \cite{Wolfire2003} thermal equilibrium curve for 8.5~kpc, obtained by assuming a constant diffuse heating rate of $22.4\times 10^{-27}$~erg~cm$^{3}$~s$^{-1}$ described in \cite{US2016}. 

The hydrodynamic simulation has been done with the code presented in \cite{US2016}, which is  in turn based on the one described in \cite{GazolKim2010}, which uses a MUSCL-type scheme (Monotone Upstream-centered Scheme for Conservation Laws) with HLL Riemann solvers \citep{Harten1983, toro_1999} . The MHD runs result from a magnetized version of the same code based on the numerical scheme described by \cite{kim1999}.  

In these simulations, the cold gas, which we define as gas with temperatures bellow the minimal temperature of the cold stable branch (278~K) determined by the cooling function described above, represents between
0.5 and 0.6 of the total mass \citep{Marco18}. This gas appears to be moderately supersonic. The corresponding distributions of the sonic Mach number have peaks at values between 3.5 and 3.7 \citep{Marco18}, which agrees with observational data presented in \cite{Heiles2003}. As expected, the magnetic Mach number distributions for the cold gas resulting from the MHD models have peaks varying over a broader range, going from 3 to 1 as $B_0$ increases \citep{Marco18}. 

Due to the permanent Fourier forcing, the models reach an stationary state \citep[see][]{Marco18}, which allows us to improve the statistics by combining data from different snapshots. In fact, for the analysis presented in the next section we use, for each simulation, data from five late snapshots separated by $\Delta t \sim 5.4 \times 10^{6}\,yr$. 
At each snapshot we select connected points with densities $n$ above three thresholds: 50, 75, and 100~cm$^{-3}$. Then we choose to include only clumps with more than $10^3$ pixels, which at the resolution of our models, correspond to equivalent spheres with a radius $\geq 1.2$~pc. The three density thresholds for the five snapshots and the size selection leave us with $\sim 300$ clumps for each model.
\subsection{An illustrative example}\label{sec:illus}
\begin{figure}
	\includegraphics[width=\columnwidth]{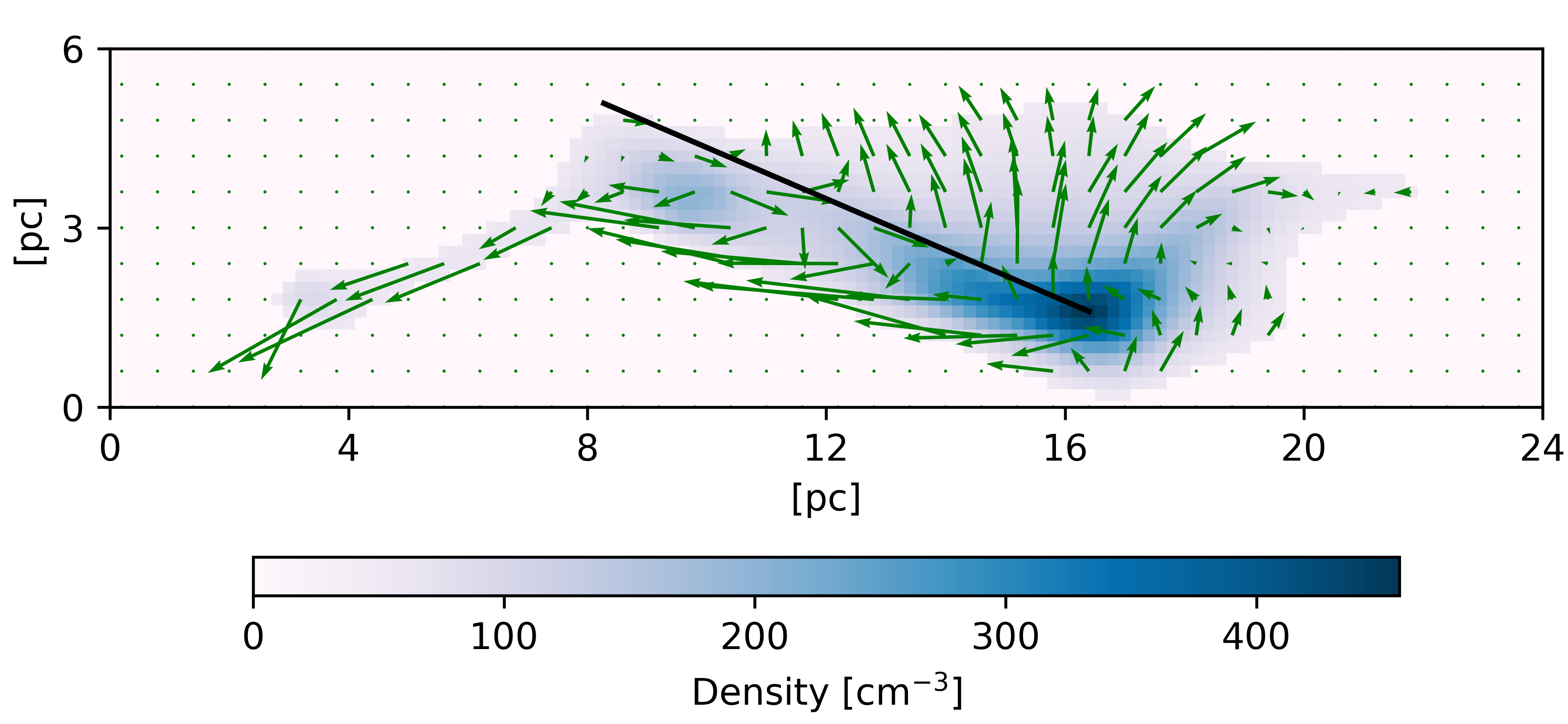}
	\includegraphics[width=\columnwidth]{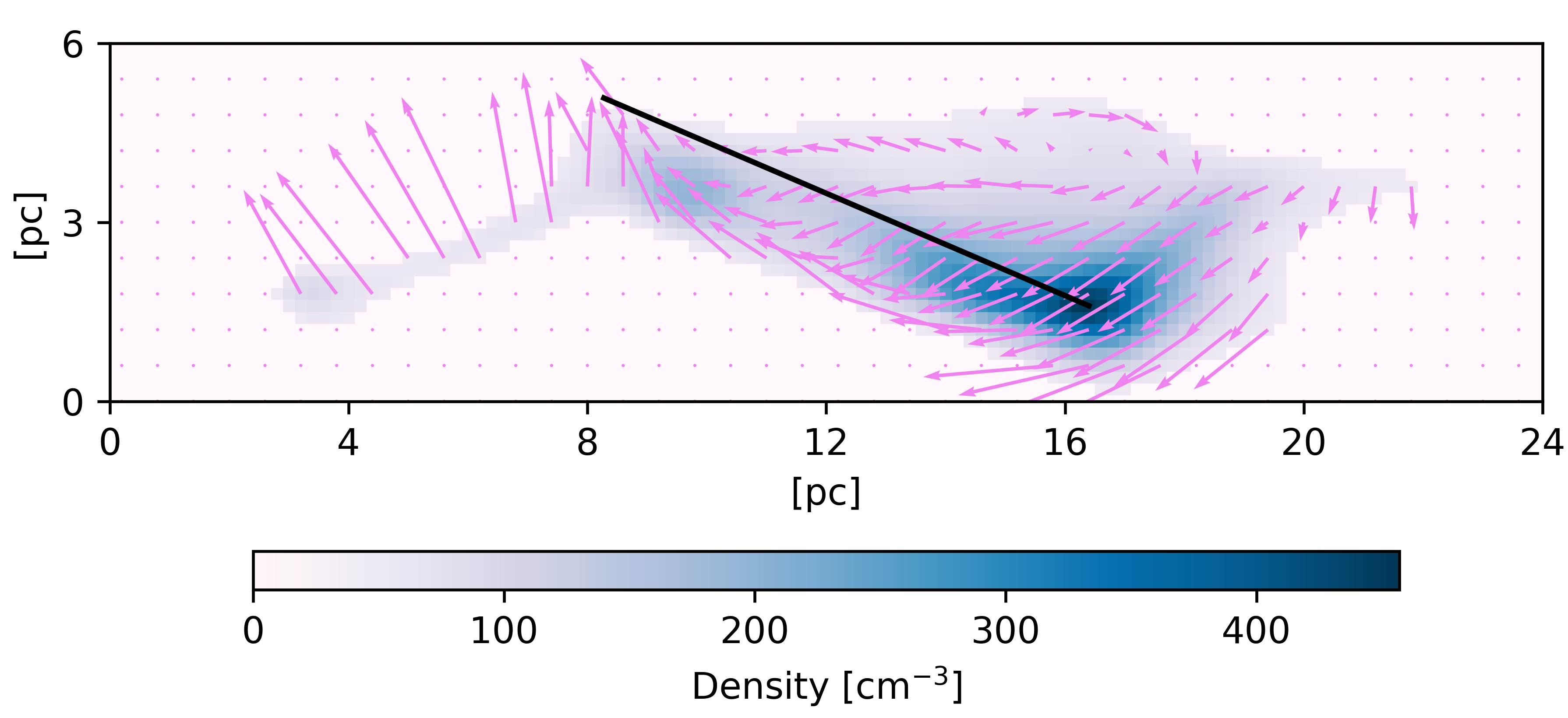}
    \caption{Two-dimensional density slice of a 3-dimensional clump selected with the lower density threshold. The superimposed vectors in each panel are the local  magnetic field (green, top) and  the local velocity field (purple, bottom), while black line is orientated in the direction of the principal axis with one extreme at the clump center of mass.}
    \label{fig:3d_clump_B01}
\end{figure}

For the sake of illustration, in this section we present an application of the analysis tools described in sections  \ref{sec:shapedesc}   and \ref{sec:dirstat} to a single randomly chosen clump. We picked-up a clump resulting from the magnetized model with the lowest initial field (B01S) and selected by using the lowest density threshold.
The direction of its principal axis as well as the
local 2-dimensional magnetic field (top) and velocity (bottom) inside the clump are shown in Fig. \ref{fig:3d_clump_B01} superimposed to a density slice. For this clump, the mean density, mean temperature, and mean magnetic field are 91~cm$^{-3}$, 74~K, and 3.7~$\mu$G, respectively. Its plasma-$\beta$ parameter is 4.4, indicating that it is dominated by thermal pressure. Concerning the shape, the asphericity $A_3$ (eqn. (\ref{eq:aspher_3d})) is 0.24 and the prolatness  $S_A$  (eqn. (\ref{eq:prolatS})) -0.12. These values indicate a moderately non spherical oblate structure, consistent with the values of its aspect ratios $\gamma=\sqrt{m_1/m_3}=0.22$ and $\beta=\sqrt{m_2/m_3}=0.81$.  From Fig. \ref{fig:3d_clump_B01} it can also be seen that it is not possible to individuate a single tendency in the relative orientation between the density structure and magnetic field or the velocity and neither in the relative orientation between these two vectors. The mean angles that we obtain from directional statistics for this clump are 0.33$\pi$ and 0.32$\pi$ for the orientation between the longest principal axis and the local magnetic field and the local velocity, respectively, and 0.26$\pi$ for the relative orientation between the local magnetic field and the local velocity. These mean angles values turn out to be highly representative for clumps resulting from  the B01S model (section \ref{sec:res_angles}). The same is true for the plasma-$\beta$ parameter (section \ref{sec:res_prop}) but is not the case for the parameters describing the clump morphology (section \ref{sec:res_shape}).   
\section{Results}\label{sec:results}
\subsection{Physical Properties of Clumps}\label{sec:res_prop}
\begin{figure}
	\includegraphics[width=\columnwidth]{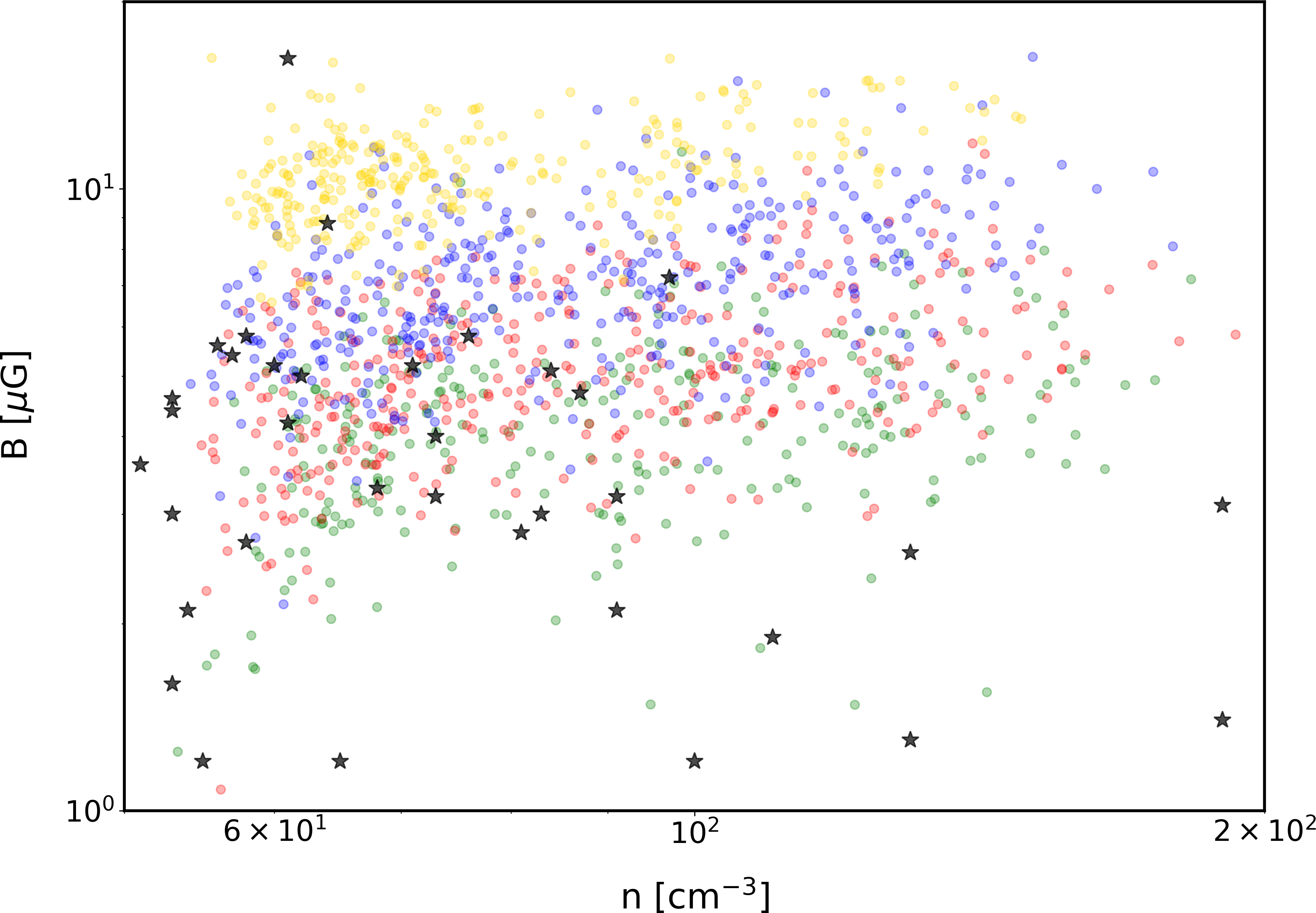}
   \caption{Magnetic field intensity $B$ vs. number density $n$ for all the selected clumps (circles) and for observational data reported by \protect\cite{HeilesTroland2004} 
   (stars). Yellow, blue, red, and green are used for clumps resulting from B20S, B10S, B05S, and B01S models, respectively.}
   \label{fig:crutcherplot}
\end{figure}

\begin{figure}
	\includegraphics[width=\columnwidth]{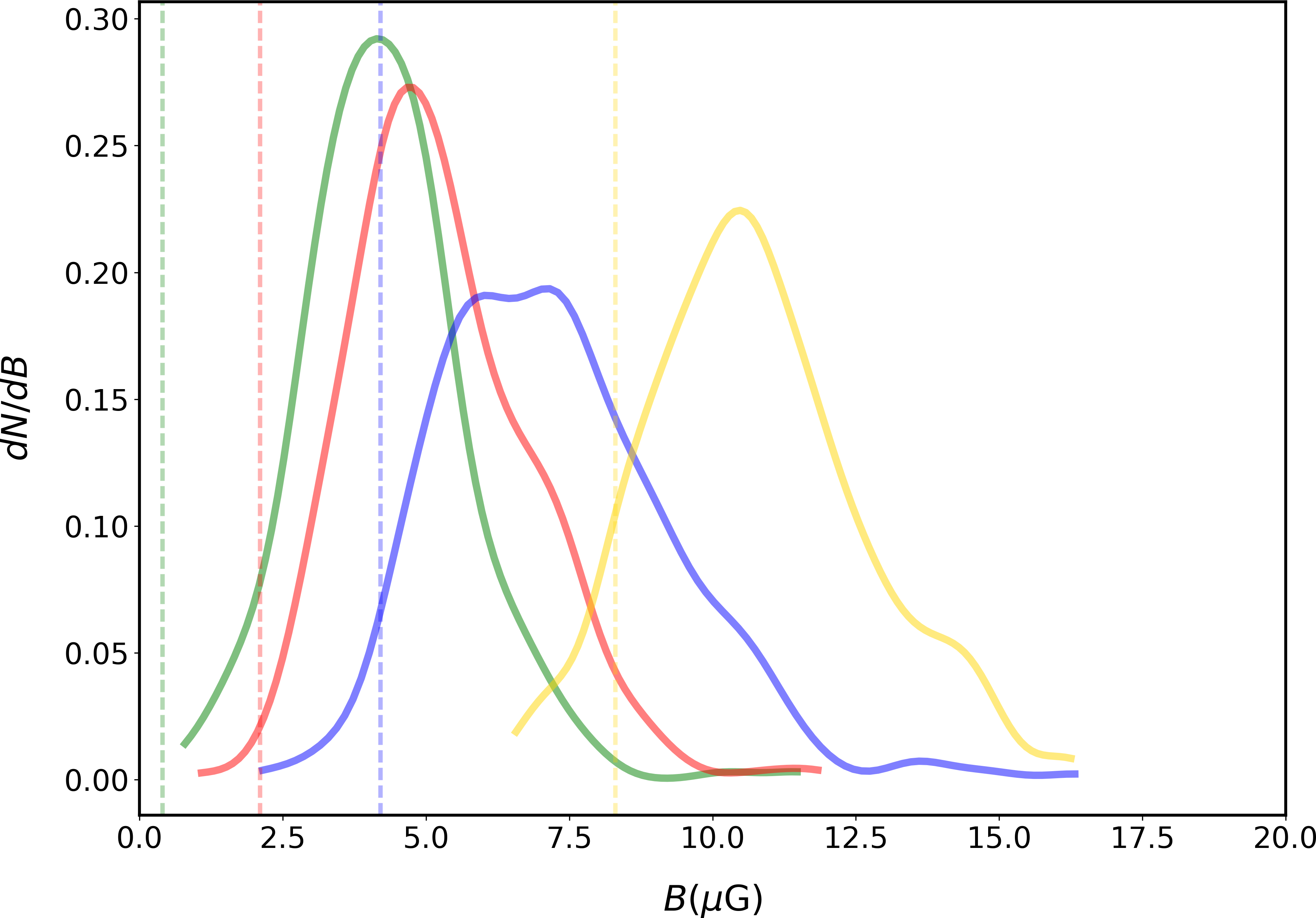}
  \caption{Mean magnetic field intensity probability density distributions from KDE. Colour code is the same as in \ref{fig:crutcherplot}. Vertical lines represent the initial magnetic field intensity $B_0$ for each model.}
    \label{fig:meanb}
\end{figure}

In Fig. \ref{fig:crutcherplot}, we show a  $B$ vs. $n$ dispersion plot for all the clumps that we take into account,  different colours correspond to different $B_0$ values:  green, red, blue, and yellow points are for $B_0=0$.4,  2.1, 4.2, and 8.3~$\mu$G, respectively. The same colour code is used for all figures in the paper.  The values of  $B$ and $n$ that we show are volume averaged quantities. The clumps in our sample have densities and magnetic field intensities comparable with those reported by observations. In fact,  black stars in Fig. \ref{fig:crutcherplot} correspond to the volume density and parallel (to the line of sight) magnetic field strength reported by \cite{HeilesTroland2004} for CNM clumps in the solar neighbourhood.  Note that, because we are considering the three dimensional magnetic field,  the values of $B$ obtained for our sample of clouds  are expected to be systematically larger than those reported by \cite{HeilesTroland2004}. From the same figure, it also can be noticed that for each $B_0$ the resulting clumps have a wide range of magnetic field intensities.  The density distribution of $B$ in clumps resulting from each model is shown in Fig. \ref{fig:meanb}, where it can be seen that there is a shift  between the distribution peak and the initial magnetic field intensity and that this shift gets relatively smaller as $B_0$ increases, as expected from the magnetic tension/pressure enhancement. It is remarkable that, despite the obvious dependence of the peak location with $B_0$,  
even for the weakly magnetized model the distribution peaks above  $3$~$\mu$G, and that for B20S the peak is in $\sim 11$~$\mu$G. 
  
\begin{figure}
	\includegraphics[width=\columnwidth]{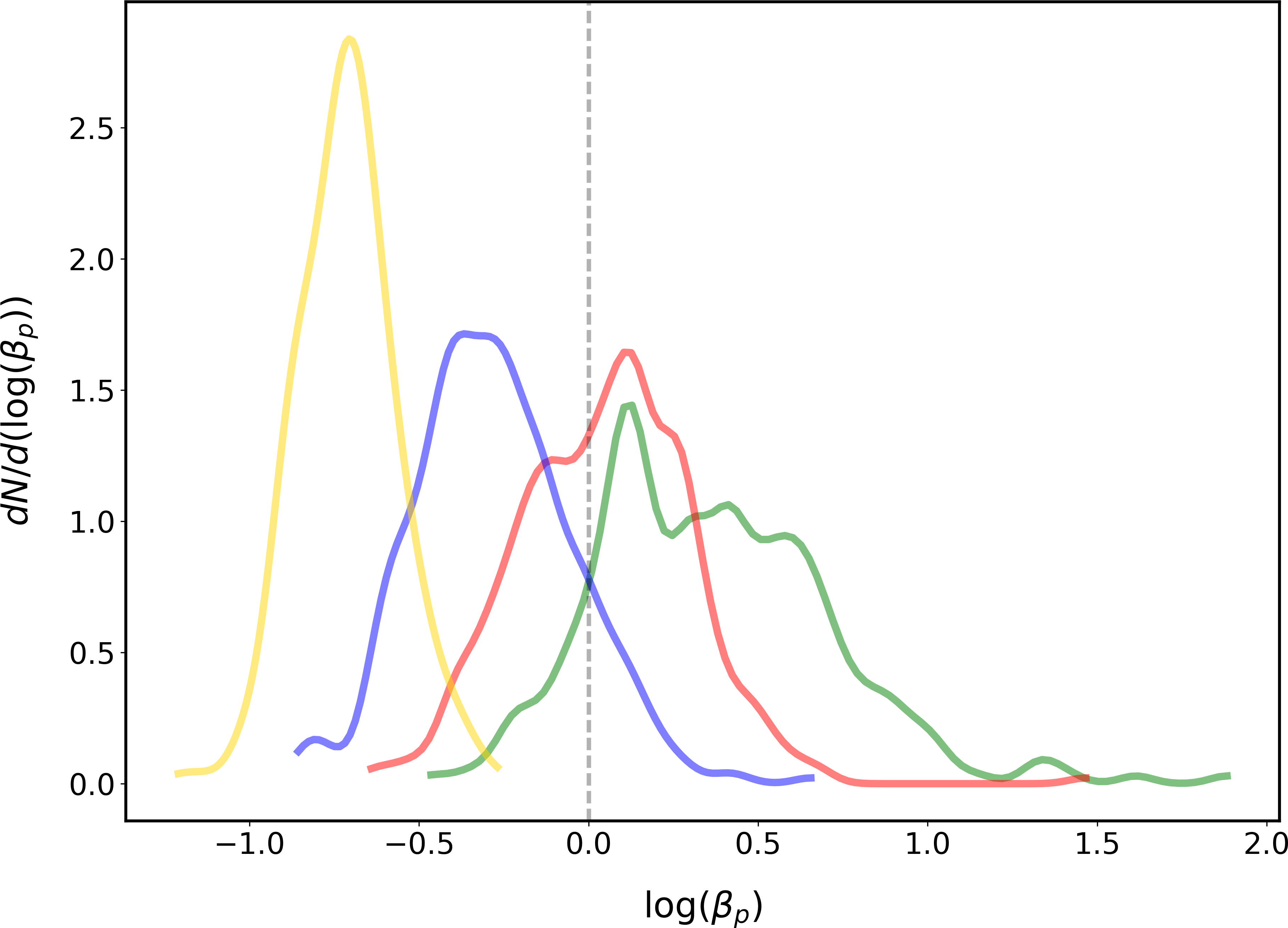}
    \caption{Mean plasma-$\beta$ parameter, $\beta_P$,  density distributions from KDE. The colour code is the same as in Fig. \ref{fig:crutcherplot}.}
    \label{fig:betas}
\end{figure}

\begin{figure}
	\includegraphics[width=\columnwidth]{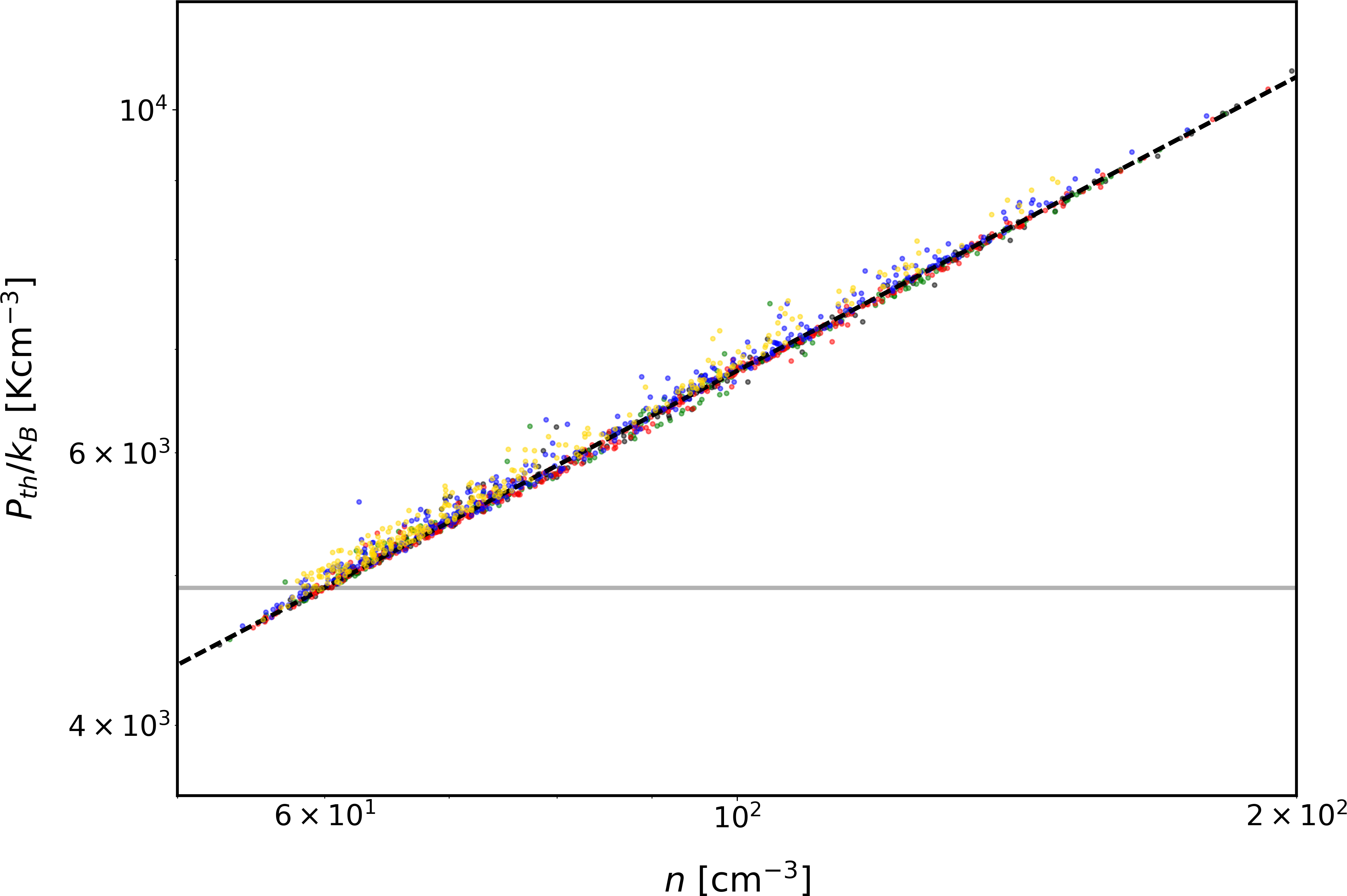}
    \caption{Thermal pressure vs. density for all the clumps in the sample. The colour code is the same as in Fig. \ref{fig:crutcherplot}, but includes black points for clumps resulting from the B00S model. Horizontal solid line is located at the maximum pressure for two phases in pressure equilibrium. Dashed line represents the thermal equilibrium curve.}
    \label{fig:thermaleq}
\end{figure}

%
Despite the relatively large magnetic field intensities, the clumps in the sample are not all dominated by magnetic pressure. The plasma-$\beta$ parameter $\beta_P=P_{th}/P_b$ ($P_{th}$ and $P_b$ are the thermal and the magnetic pressure, respectively), whose distributions are displayed in Fig. \ref{fig:betas},  is preferentially $<1$ only for clumps resulting from simulations with large magnetic fields, B10S and B20S. For B05S, a balance between the thermal and the magnetic pressure is attained by an important fraction of clumps, and the $\beta_P$ distribution shows no clear preference for $\beta_P<1$ nor  $\beta_P>1$. Clumps resulting from the MHD model with the lowest $B_0$, B01S, are preferentially dominated by thermal pressure. As can be noticed in Fig. \ref{fig:thermaleq}, all the selected clumps, regardless of the model from which they result, are in thermal equilibrium at thermal pressures above the maximum pressure for two phases existing in pressure equilibrium. Note that the figures described to this point, as well as figures in the rest of the paper, we do not make the distinction between clumps resulting from different density thresholds.   In fact we do not observe any systematic behavior, other than in quantities which directly involve $n$ or $P_{th}$, depending on the density threshold.

\begin{figure}
	\includegraphics[width=\columnwidth]{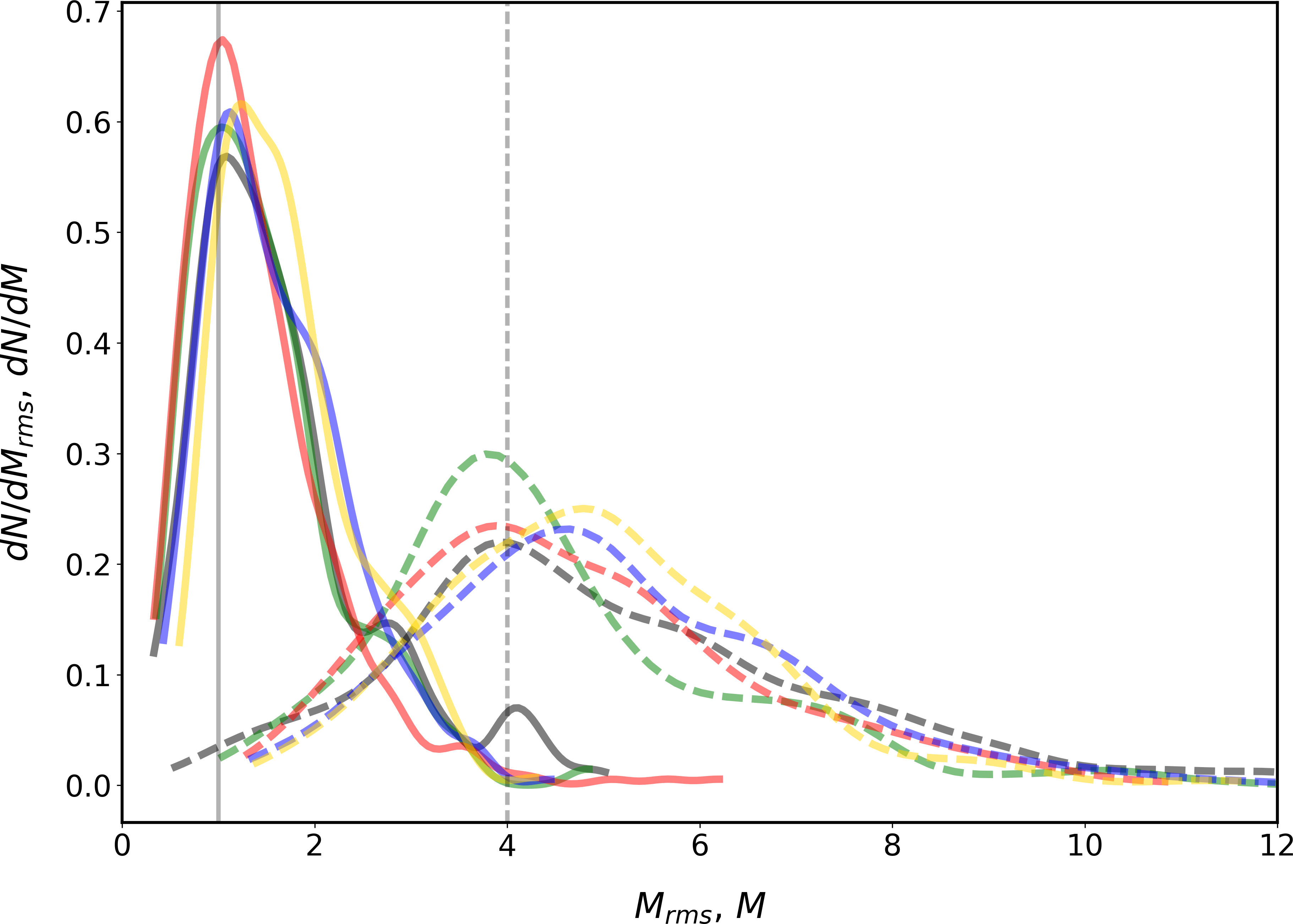}
    \caption{Probability density from KDE for the internal sonic Mach number, $M_{rms}$ (solid lines),  and the total sonic Mach number $M$ (dashed lines). The colour  code is  the same as in Fig. \ref{fig:thermaleq}. Vertical gray lines are placed for reference at 1 and 4.}
    \label{fig:Mach}
\end{figure}

In Fig. \ref{fig:Mach}, the  PDF for two different estimations of the sonic Mach number inside the clumps are shown.
The internal turbulent Mach number (solid lines) is defined as $M_{rms}=\langle  v_{rms}/c_s\rangle$, where $ v_{rms}=|\mathbf{v}-\mathbf{\bar v}|$, $c_s$ is the local sound speed, and  $\langle \rangle$ denotes the average over all the pixels within each clump. The clumps in our sample have internal motions that are preferentially trans-sonic or slightly supersonic.
On the other hand, the total sonic Mach number $M=\langle v/c_s \rangle $ (dashed lines), which includes the bulk motion of the clumps, has clearly larger  preferential values, around $M=4$. This implies that the supersonic nature of the CNM structures resulting from our simulations is mainly due to the relative motions of the clumps with respect to the ambient medium.


%

\begin{figure}
	\includegraphics[width=\columnwidth]{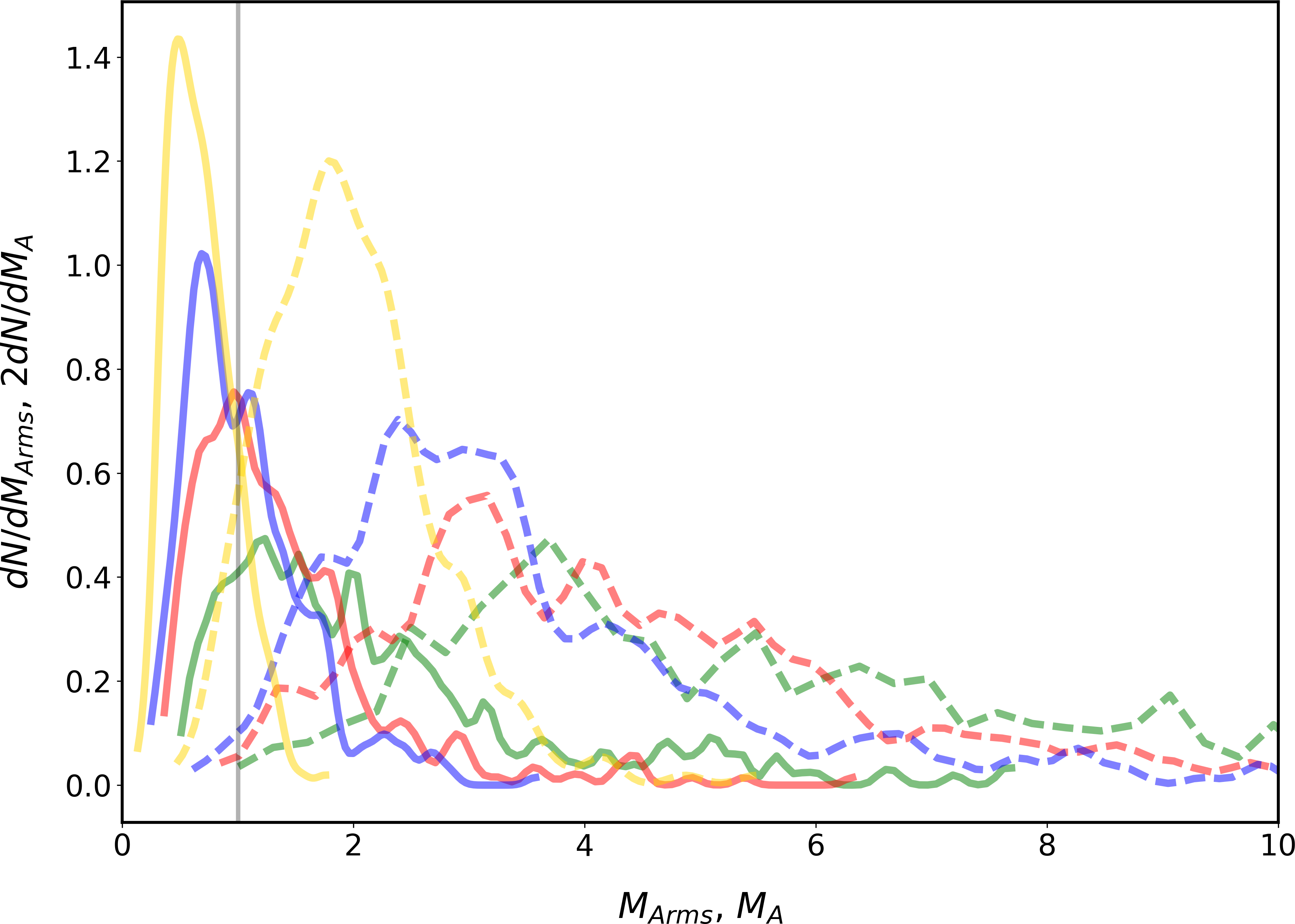}
    \caption{Probability density from KDE for the internal Alfv\'en Mach number, $M_{Arms}$ (solid lines), and the total Alv\'en Mach number $M_A$ (dashed lines). The vertical gray vertical line  is placed at 1  for reference, and the colour  code is  the same as in Fig. \ref{fig:crutcherplot}}
    \label{fig:MachAs}
\end{figure}

As for the sonic case, we display two different estimations for the Alv\'en Mach number (Fig. \ref{fig:MachAs}). The purely internal Alfv\'en Mach number, $M_{Arms}$, computed by using  $v_{rms}$, (solid lines) shows that
clumps resulting from B20S are almost exclusively sub-Alfv\'enic and have a PDF with a well defined single peak , while
those resulting from B01S are almost exclusively super-Alfv\'enic with an extended distribution having significant values of up to  $M_{Arms}\sim 3$. Clumps resulting from models with intermediate values of $B_0$ show that the broadening of the distribution is progressive as $B_0$ decreases. For the B10S model, the distribution has a clear peak in the sub-Alfv\'enic region and a lower although noticeable peak in the super-Alfv\'enic zone, while in the  B05S case, the distribution peaks at 1. On the other hand, the distributions of the total Alv\'en Mach number $M_A=\langle v/v_A \rangle $ (Fig. \ref{fig:MachAs}, dashed lines), which includes the bulk motions of the cloud, peak at values $>1$ and $<5$ for all the models and, as expected, its width increases as $B_0$ decreases. 
%
\subsection{Morphology}\label{sec:res_shape}
\begin{figure}
	\includegraphics[width=\columnwidth]{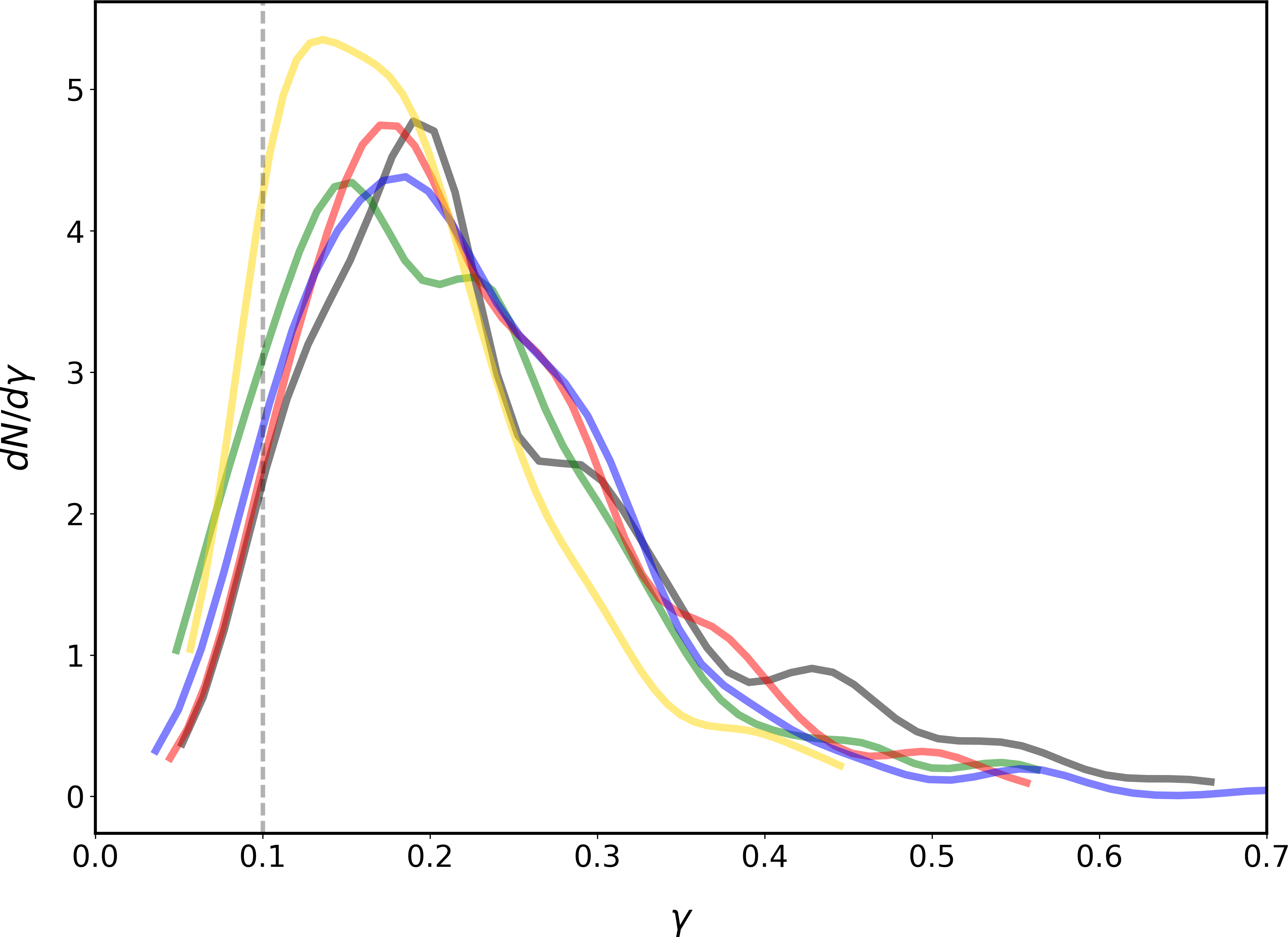}
    \caption{Density distribution function for the largest to smallest semi-axes length ratio $\gamma$. The colour code is the same as in Fig. \ref{fig:thermaleq}}
    \label{fig:aspectratio}
\end{figure}

\begin{figure}
	\includegraphics[width=\columnwidth]{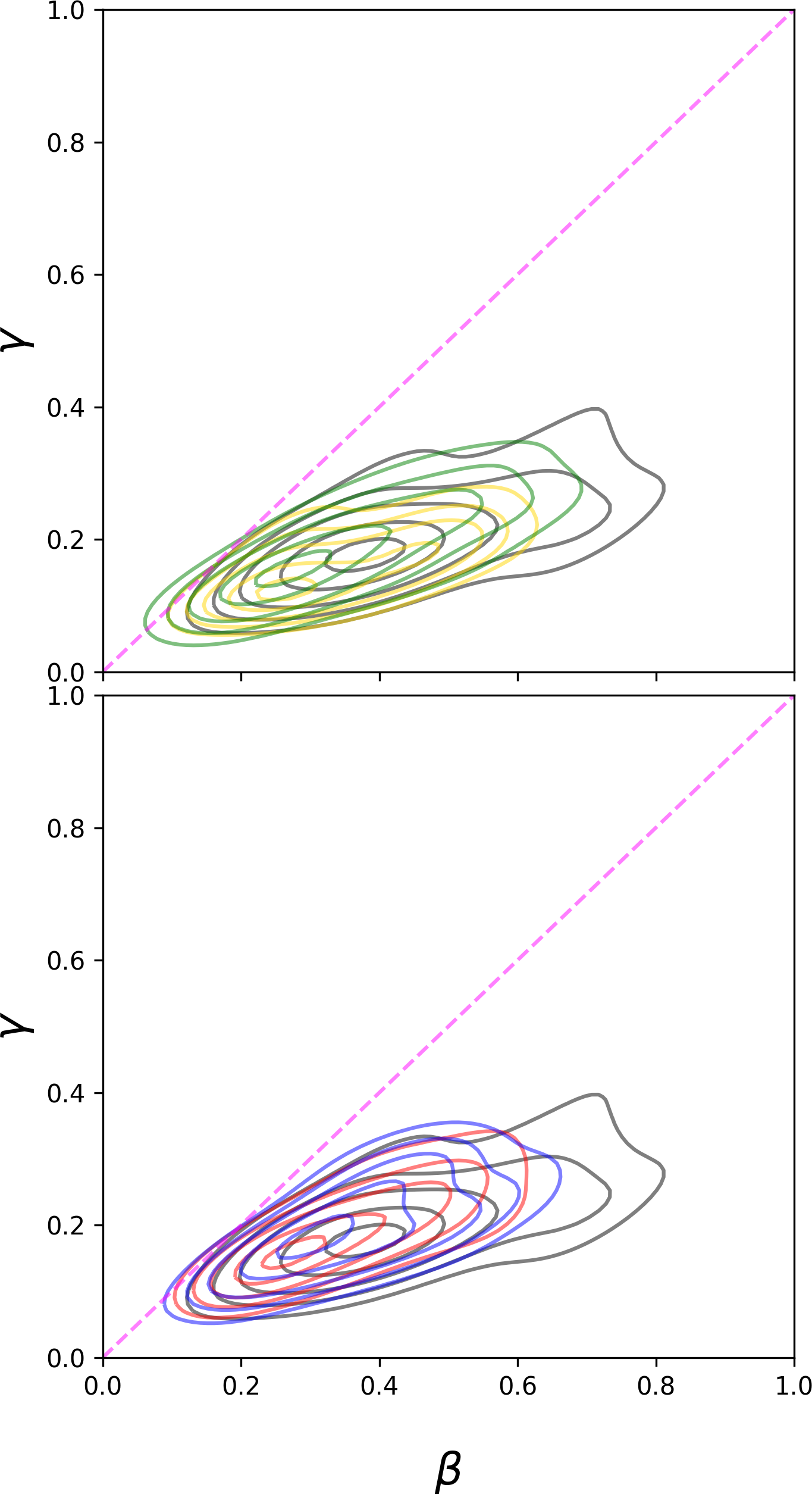}
    \caption{2d-KDE contour plots for the distribution of aspect ratios $\gamma$ vs. $\beta$, for  B00S, B01S, and, B20S (top panel) and for B00S, B05S, and B10S (bottom panel). The colour code is the same as in Fig. \ref{fig:thermaleq}. For reference, the $\gamma=\beta$ line (magenta) is also included.}
    \label{fig:aspectratio2d}
\end{figure}

The distribution of the ratio between the lengths of the smallest and the largest semi-axes, $\gamma$, is plotted in Fig. \ref{fig:aspectratio}. It can be noticed that for the magnetized runs the PDF peaks at smaller values, indicating a preference for more anisotropic clumps. When comparing the non magnetized model (black line) with the one with the largest $B_0$ (yellow line), it is clear that in the former case the PDF is more populated for $\gamma\gtrsim 0.25$ and reaches larger values, but the distinction between the purely hydrodynamic model and and magnetized models other than B20S is not so clear. When the probability of having $\gamma\leq 0.25$, $\mathcal{P}_\gamma$  is computed for each model (third column in Table \ref{tab:shape}), the larger difference is between B00S and B20S, but MHD models appear to have larger values of  $\mathcal{P}_\gamma$. For the intermediate $B_0$ values, however, the difference with the hydrodynamic model is small. The same behaviour, concerning the extreme cases, i.e. B00S and B20S,  is observed when the two dimensional KDs for  $\gamma$ vs. $\beta$ are examined (Fig. \ref{fig:aspectratio2d}), it can be seen that in the non magnetic case clumps tend to be at regions with less elongated geometries. Note that in this plot, the right upper corner corresponds to spherical clumps, while clumps near the $\beta=1$ limit and low $\gamma$ are oblate (disk-like). On the other hand, objects near the $\gamma=\beta$ line are prolate (rod-like). Following a point near the upper right corner and below the line $\gamma=\beta$, towards the lower left corner remaining just below this line, describes prolate objects changing from a slightly elongated clump to a filament. 
In fact,  for the B20S model the contours  exhibit an enhanced concentration towards low $\beta$ and low $\gamma$ which places it near the $\gamma=\beta$ line in the filamentary region. This concentration  is not present in contours corresponding to  the purely hydrodynamic model (gray lines), which show a peak at larger $\gamma$ and $\beta$ values and extend towards larger $\beta$ values. Note that contours are placed at fixed values relative to the maximum reached by each distribution. The $\beta$ shift in peak location as well as the more concentrated distributions are also distinguishables when comparing the hydrodynamic model with the other three MHD models (green, red, and blue lines) but it is not possible to distinguish a trend in the contour distribution as $B_0$ changes. This fact motivates us to explore other shape descriptions with the aim of finding more quantitative trends on clump shape distributions with
$B_0$.   


\begin{table}
\caption{Morphology probabilities}
\label{tab:shape}
\begin{tabular}{lccccccc}
\hline
Model &$B_0$    & $\mathcal{P}_{\gamma}$$^\textbf{a}$ & $\mathcal{P}_{A_3L}$$^\textbf{b}$& ${\mathcal{P}}_{A_3H}$$^\textbf{c}$ &
$\mathcal{P}_{SO}$$^\textbf{d}$ & $\mathcal{P}_{SPL}$$^\textbf{e}$ &$\mathcal{P}_{SPH}$$^\textbf{f}$
\\
& ($\mu$G)& &&&&&\\
\hline
B00S & -  & 0.63 & 0.22 & 0.13 & 0.14 & 0.24 & 0.14 \\
B01S& 0.4 & 0.71 & 0.12 & 0.24 & 0.08 & 0.15 & 0.26\\
B05S& 2.1 & 0.67 & 0.12 & 0.22 & 0.08 & 0.14 & 0.24\\
B10S& 4.2 & 0.68 & 0.09 & 0.20 & 0.06 & 0.13 & 0.21\\
B20S& 8.3 & 0.82 & 0.07 & 0.27 & 0.04 & 0.10 & 0.30\\
\hline
\end{tabular}
\begin{itemize}
    \item[$^\textbf{a}$]\tiny{$\gamma \leq 0.25$}
    \item[$^\textbf{b}$]\tiny{$A_3\leq 0.25$}
    \item[$^\textbf{c}$]\tiny{$0.75\leq A_3\leq 1$}
    \item[$^\textbf{d}$]\tiny{$S_A<0$}
    \item[$^\textbf{e}$]\tiny{$0 < S_A < 0.25$}
    \item[$^\textbf{f}$]\tiny{$1.25 < S_A < 2$}
\end{itemize}
\end{table}

\begin{figure}
	\includegraphics[width=\columnwidth]{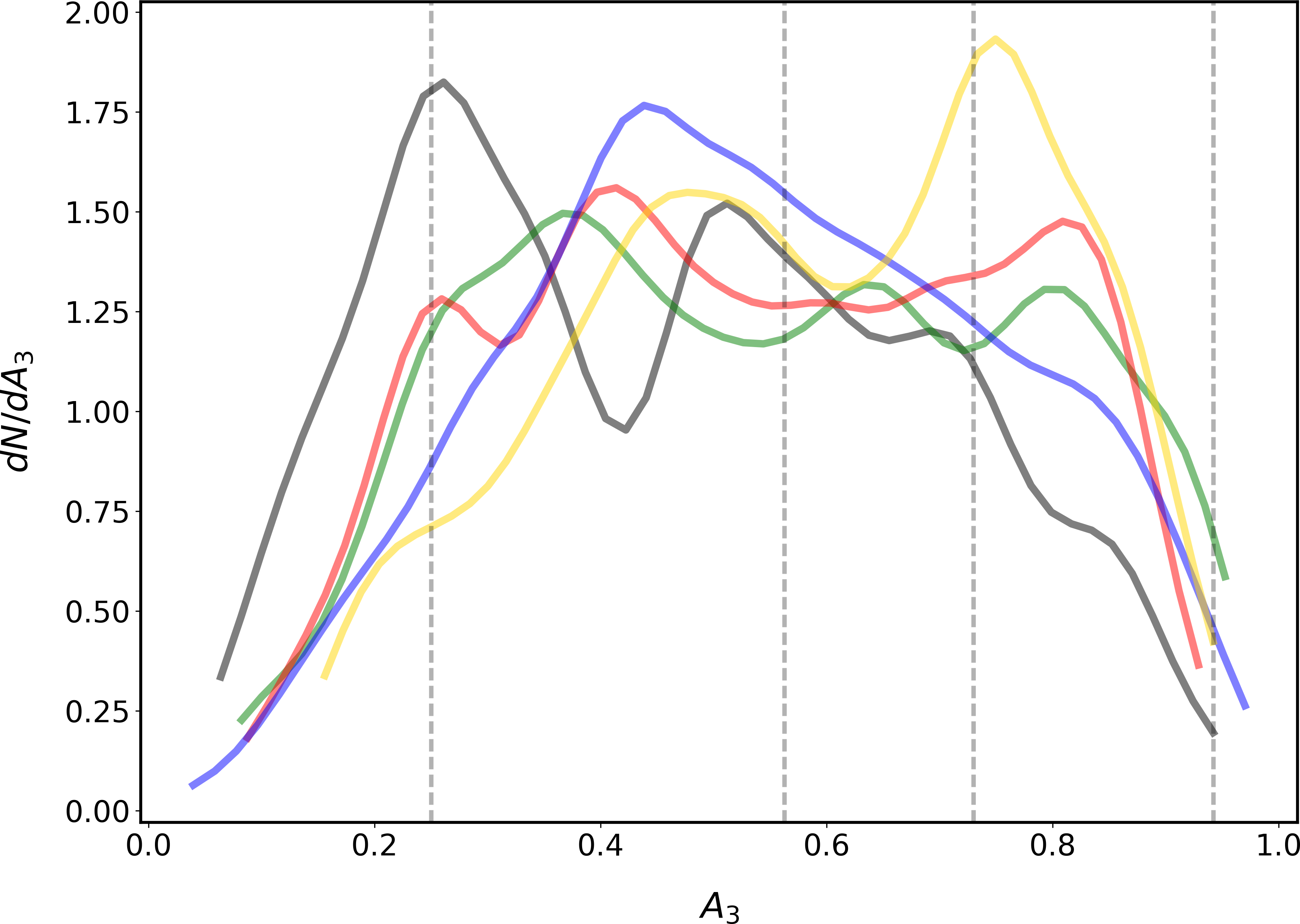}
    \caption{Density distributions for the asphericity $A_3$. The colour code is the same as in Fig. \ref{fig:thermaleq}. The vertical dashed lines are placed for reference at $A_3= 0.25$, 0.5625, 0.73, and 0.94; which result, for instance, from $(m_1, m_2, m_3) = (0,1,1)$, $(1,1,10)$, $(1, 10, 100)$, and $(1,1,100)$, respectively.}
    \label{fig:asphericity3d}
\end{figure}

The density distributions for the three-dimensional asphericity, $A_3$ (eqn. (\ref{eq:aspher_3d})), are plotted in Fig. \ref{fig:asphericity3d}, where it can be seen that the tendency of clumps resulting from the non-magnetized run to be more spherical is recovered. However, three additional facts can be noticed from this figure. First, even for the MHD model with the lowest $B_0$, B01S, the presence of the magnetic field produces less spherical structures in both ways: the low asphericity region becomes less populated and the high asphericity region becomes more populated for magnetized models. Second, the scarcity of nearly spherical magnetized clumps is more pronounced than the scarcity of highly non spherical clumps in the non-magnetized case. And third, the three-dimensional asphericity distribution does not seem to  follow a simple trend as $B_0$ increases. We recover these trends
when we look at the probability of having $A_3\leq 0.25$, $\mathcal{P}_{A_3L}$ (fourth column in Table \ref{tab:shape}), and the probability of having  $A_3\geq 0.75$, $\mathcal{P}_{A_3H}$ (fifth column in Table \ref{tab:shape}). From the values of $\mathcal{P}_{A_3L}$ and $\mathcal{P}_{A_3H}$ it can be seen that with $A_3$ the morphological differences between clumps from magnetized models and clumps resulting from a purely hydrodynamic  model are more notorious. 

As can be seen from the definition of $A_3$ (eq. (\ref{eq:aspher_3d})), the same value of the asphericity can be obtained for different sets of eigenvalues, i.e. for different shapes. For example, the asphericity $A_3 \sim 0.25$ can be obtained with different sets of eigenvalues, two of them are ($m_1 = 0, m_2 = 1, m_3 = 1$) and ($m_1 = 5, m_2 = 5, m_3 = 20$). The first set corresponds to a sheet-like surface while the second one represents a rectangular box. For the intended purposes of this paper those two objects are entirely different and therefore, in order to further classify them, a refinement is needed. This refinement is achieved with the measure of the object's prolatness, which distinguishes between elongated and flattened non-spherical objects. 

\begin{figure}
	\includegraphics[width=\columnwidth]{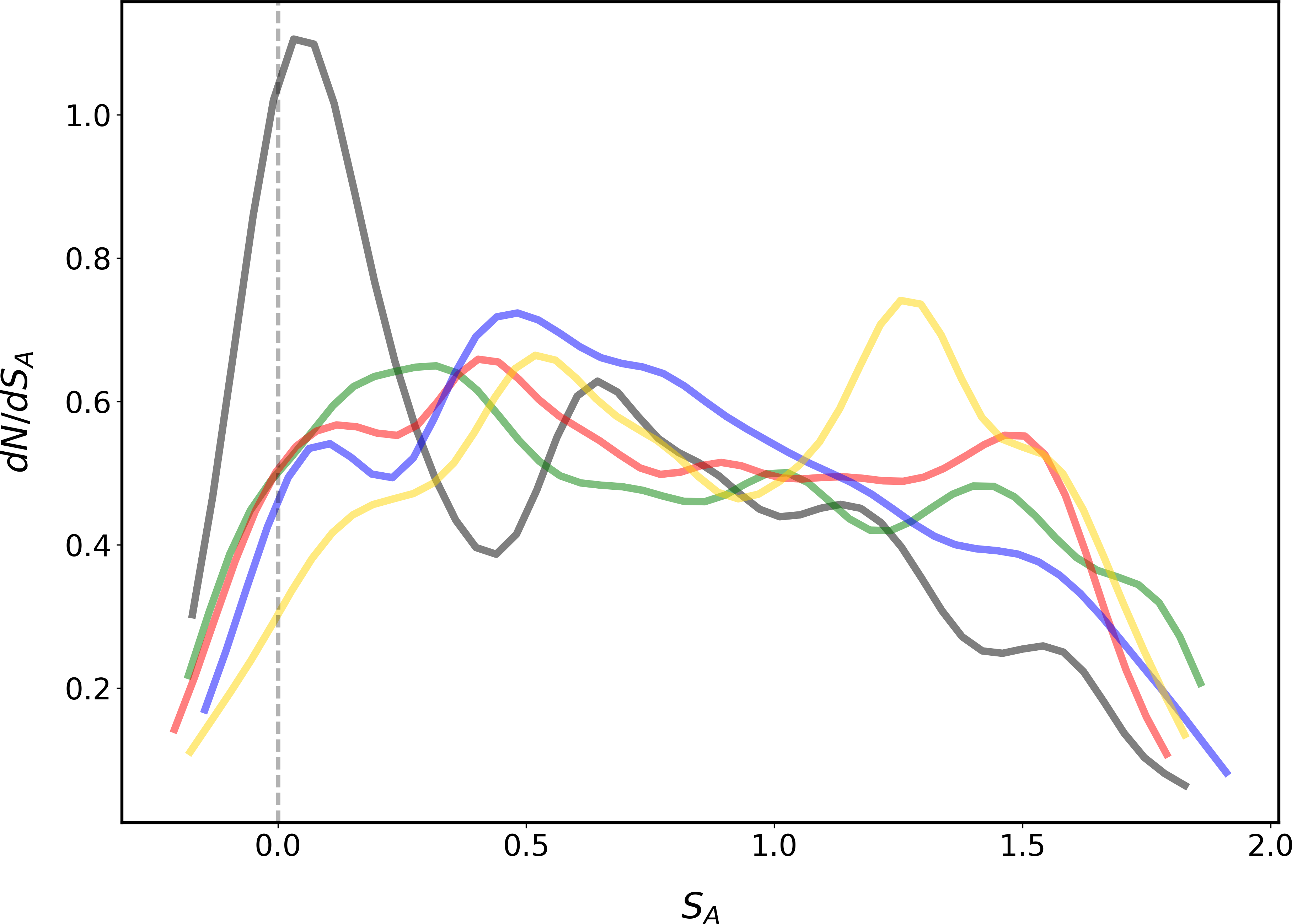}
    \caption{Density distributions for the prolatness $S_A$. The colour code is the same as in Fig. \ref{fig:thermaleq}.}
    \label{fig:prolatness}
\end{figure}

The evaluation of the prolatness $S_A$ (eqn. (\ref{eq:prolatS})), displayed in Fig. \ref{fig:prolatness},  shows that the clumps resulting from our models are mainly prolate, i.e.  are more prone to have a filament-like geometry rather than a sheet-like geometry. This is true both, in the non magnetic case as well as in the magnetic cases. The probability of having $S_A<0$, $\mathcal{P}_{SO}$ (sixth column in Table \ref{tab:shape}), is however larger for the hydrodynamic model and it decreases with $B_0$. The density distributions show also that the B00S model presents a peak just above $S_A=0$, which reflects the fact that these clumps are less anisotropic. We recover that behavior when looking at the probability of having $0<S_A<0.25$, $\mathcal{P}_{SPL}$ (seventh column in Table \ref{tab:shape}), but from this quantity we can also see that the probability of having slightly prolate clumps decreases as $B_0$ increases. In fact, the region $0<S_A\lesssim 0.25$ seems to attain lower densities as $B_0$ increases. On the other side, for $S_A\gtrsim 1.25$ all the magnetized models develop a more populated PDF but no systematic trend can be distinguished with  $B_0$. These trends can also be seen in the probability of having $1.25<S_A< 2$ $\mathcal{P}_{SPH}$ (eighth column in Table \ref{tab:shape}).

\subsection{Relative Alignments}\label{sec:res_angles}
\begin{figure}
    \includegraphics[width=\columnwidth]{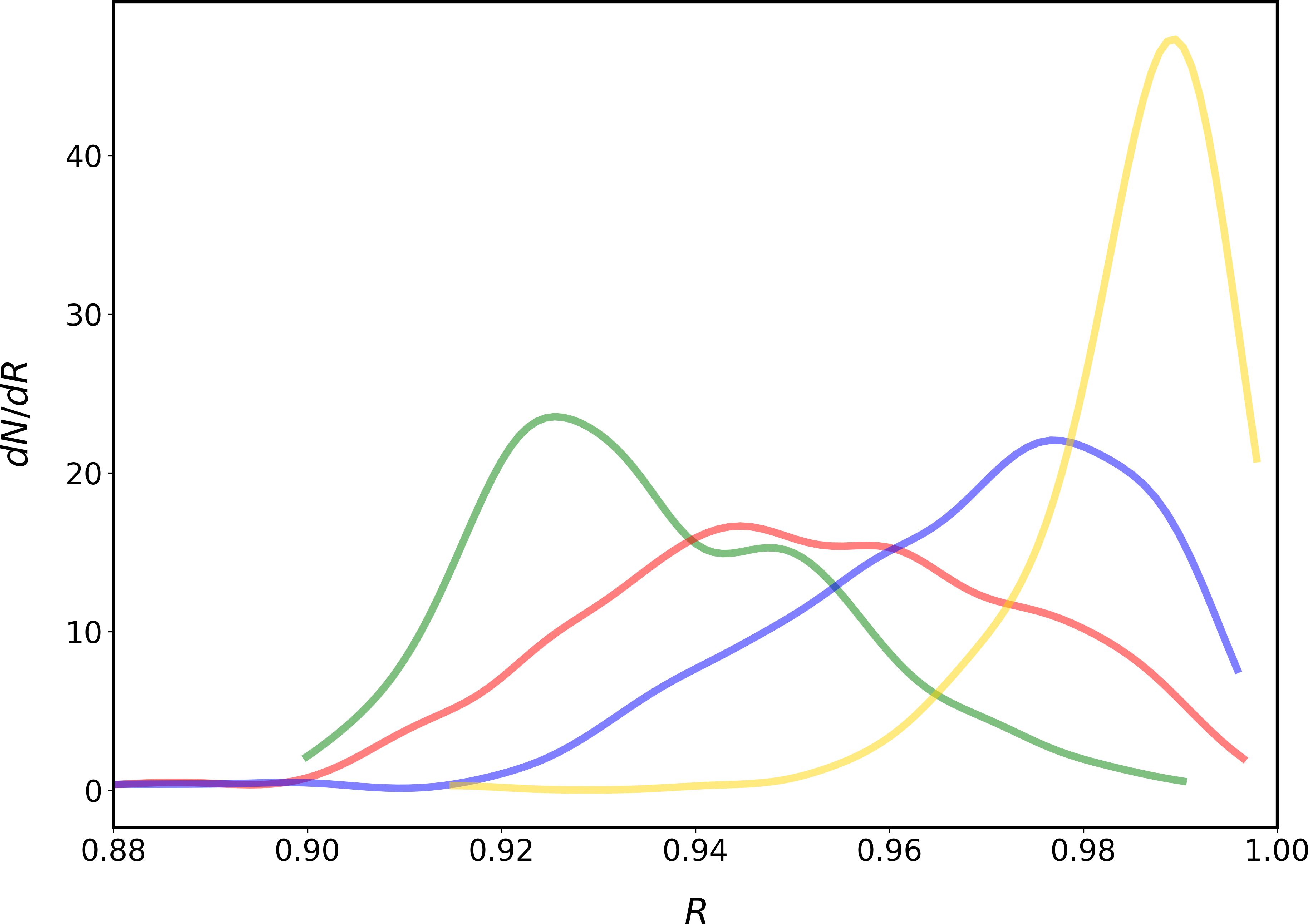}
    \caption{Density distribution functions of the distribution's center vector $\overline{R}$ for the angles between the largest principal axis and the magnetic field of the clumps. The colour code is the same as in Fig. \ref{fig:crutcherplot}}
    \label{fig:KDE_R_B}
\end{figure}

Applying the statistical tools described in section \ref{sec:dirstat} and using the KDE method described in section \ref{sec:kdes}, we obtained the PDF of the angle between the largest principal axis and the local magnetic field and the local velocity, $\overline{\theta}_B$ and $\overline{\theta}_V$, respectively; as well as the angle between the local magnetic field and the local velocity  $\overline{\theta}_{VB}$. Note that in this analysis we do not take into account the direction of the vector quantities $\vec{B}$ and $\vec{v}$, we consider only its orientation, i.e. we do not distinguish between parallel and anti-parallel or between perpendicular and anti-perpendicular relative orientations, thus the angle distributions lie in the $[0, \pi/2]$ interval. The use of directional statistics has been possible due to the high values of almost all the $\overline{R}$ measured. The most diverse values of $\overline{R}$ come from the $\theta_{B}$ angles, nevertheless all of them are quite close to unity, as can be seen in Fig. \ref{fig:KDE_R_B}. The $\overline{R}$ distributions for $\theta_{V}$ and $\theta_{VB}$ are decisively close to 1 and therefore are not shown. 

\begin{table}
\caption{Relative orientation probabilities}
\label{tab:angles}
\begin{tabular}{lcccccc}
\hline
Model &$B_0$    & $\mathcal{P}_{B}$$^\textbf{a}$ & $\mathcal{P}_{V}$$^\textbf{b}$& 
$\mathcal{P}_{VBL}$$^\textbf{c}$ & $\mathcal{P}_{VBH}$$^\textbf{d}$ 
\\
& ($\mu$G)& &&&&\\
\hline
B00S& -   & -    & 0.73 &  -   &  -   \\
B01S& 0.4 & 0.68 & 0.77 & 0.80 & 0.08  \\
B05S& 2.1 & 0.90 & 0.78 & 0.64 & 0.21  \\
B10S& 4.2 & 0.90 & 0.71 & 0.57 & 0.27  \\
B20S& 8.3 & 0.98 & 0.73 & 0.44 & 0.36  \\
\hline
\end{tabular}
\begin{itemize}
    \item[$^\textbf{a}$]\tiny{$ \pi/4 \leq \overline{\theta}_B \leq \pi/2$}
    \item[$^\textbf{b}$]\tiny{$ \pi/4 \leq \overline{\theta}_V  <   \pi/2$}
    \item[$^\textbf{c}$]\tiny{$ \pi/4 \leq \overline{\theta}_{VB} <   3\pi/8$}
    \item[$^\textbf{d}$]\tiny{$3\pi/8 \leq \overline{\theta}_{VB} \leq \pi/2$}
\end{itemize}
\end{table}

\begin{figure}
    \includegraphics[width=\columnwidth]{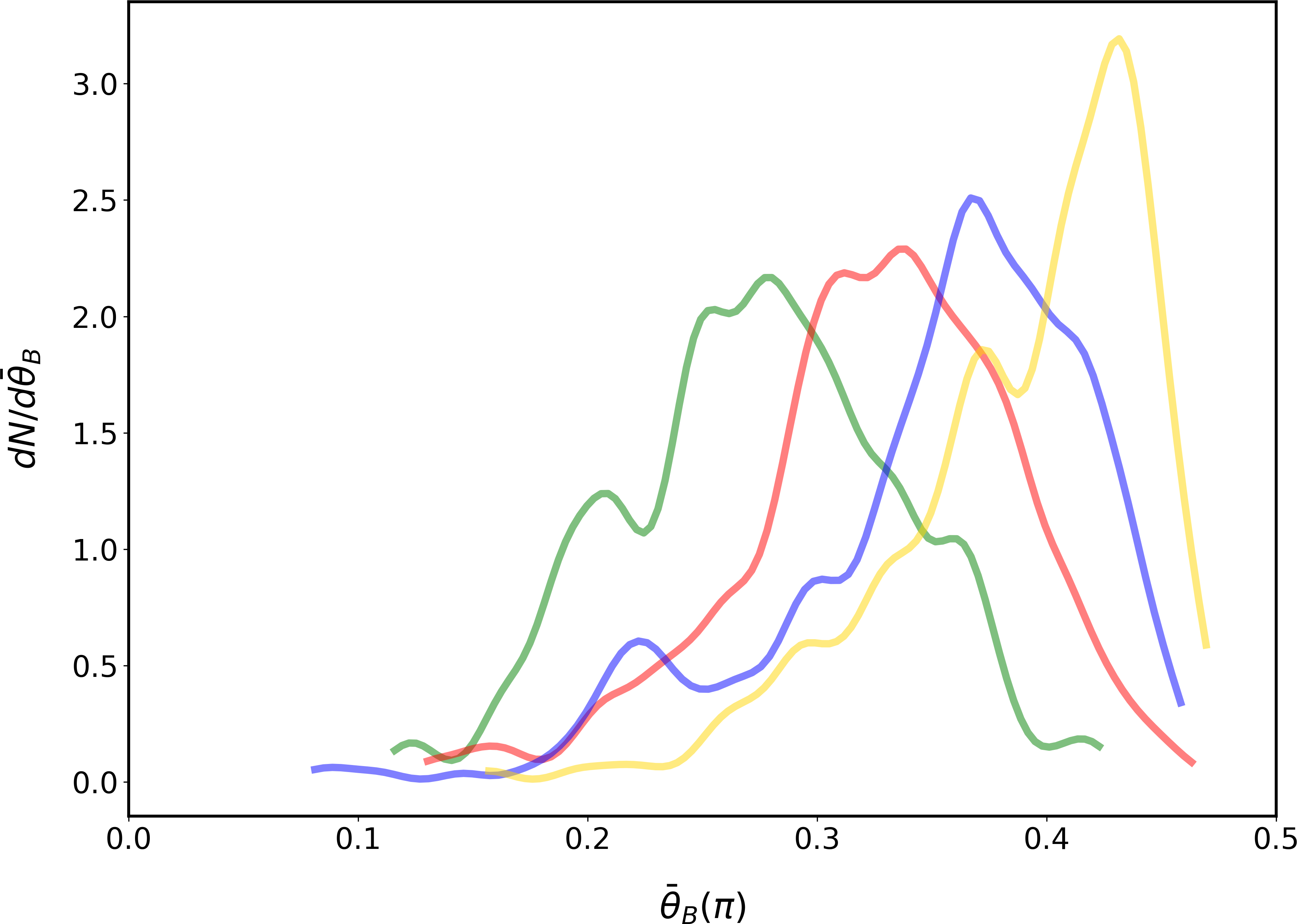}
    \caption{Density distribution functions of the mean angle $\overline{\theta}_B$ between the largest principal axis  and the local magnetic field. The colour code is the same as in Fig. \ref{fig:crutcherplot}}
    \label{fig:KDE_TH_B}
\end{figure}

The distributions of $\overline{\theta}_B$ are plotted in Fig. \ref{fig:KDE_TH_B}. It can be seen that every distribution has a different mode and a different width. They are narrower and closer to $\pi/2$ as the mean magnetic field increases. 
An angle $\overline{\theta}_B = \pi/2$ indicates that the clumps are elongated in a direction perpendicular to the internal magnetic field of the clump.  In our sample, the largest share of the clumps have angles $\overline{\theta}_B > \pi/4$ with a tendency to become closer to $\pi/2$ as $B_0$ increases. The probability of having $\pi/4\leq\overline{\theta}_B\leq\pi/2$, $\mathcal{P}_{B}$ (third column in Table \ref{tab:angles}), goes in fact from 0.68 for B01S to 0.98 for B20S. 
The relationship between this result, obtained for the averaged clump orientation,  with results gotten  by using the local clump orientation will be discussed in section \ref{secs:disc_angles}.  

\begin{figure}
    \includegraphics[width=\columnwidth]{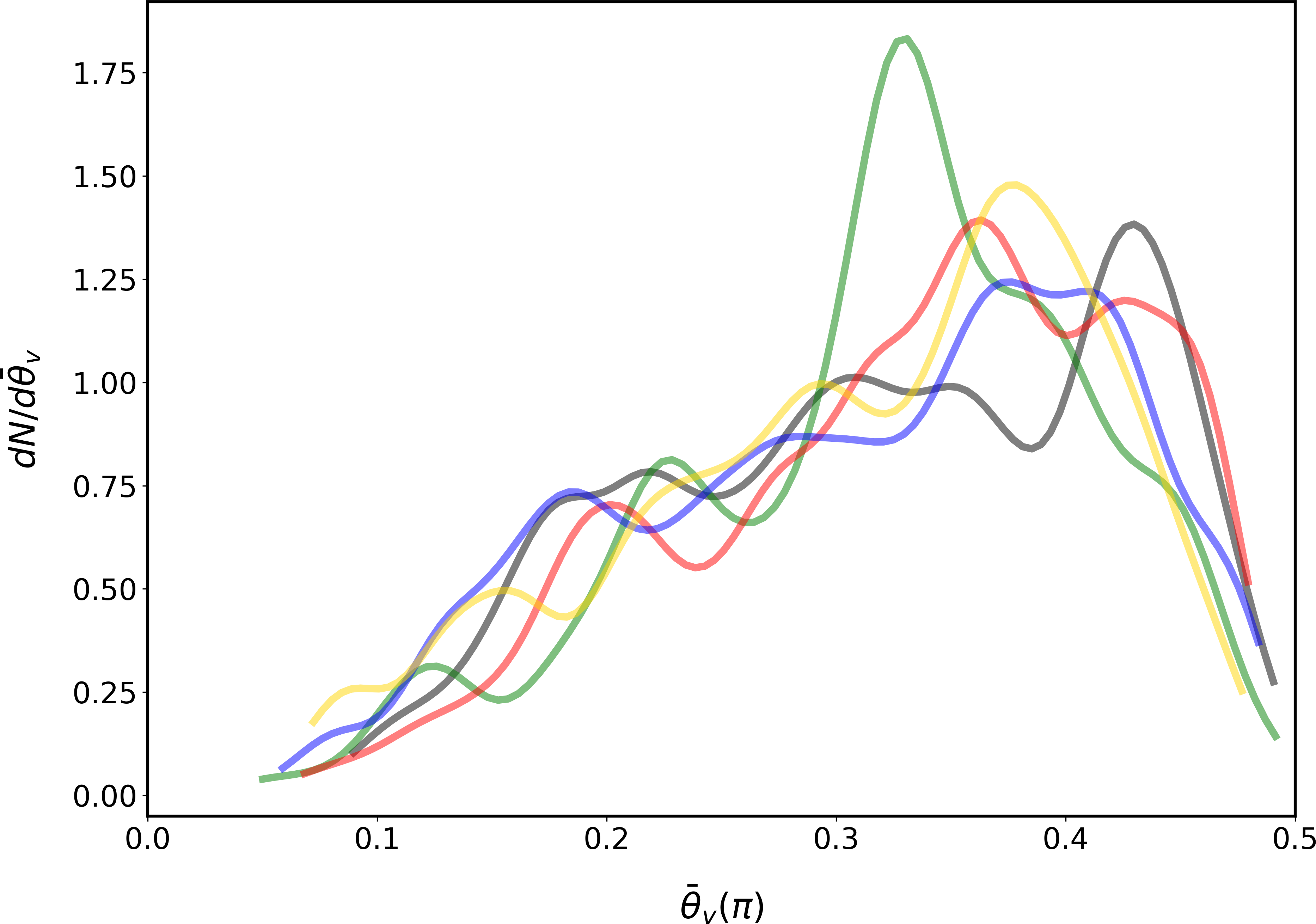}
    \caption{Density distribution functions of the mean angle $\overline{\theta}_V$ between the largest principal axis and the total local velocity. The colour code is the same as in Fig. \ref{fig:thermaleq}}
    \label{fig:KDE_TH_V}
\end{figure}

\begin{figure}
    \includegraphics[width=\columnwidth]{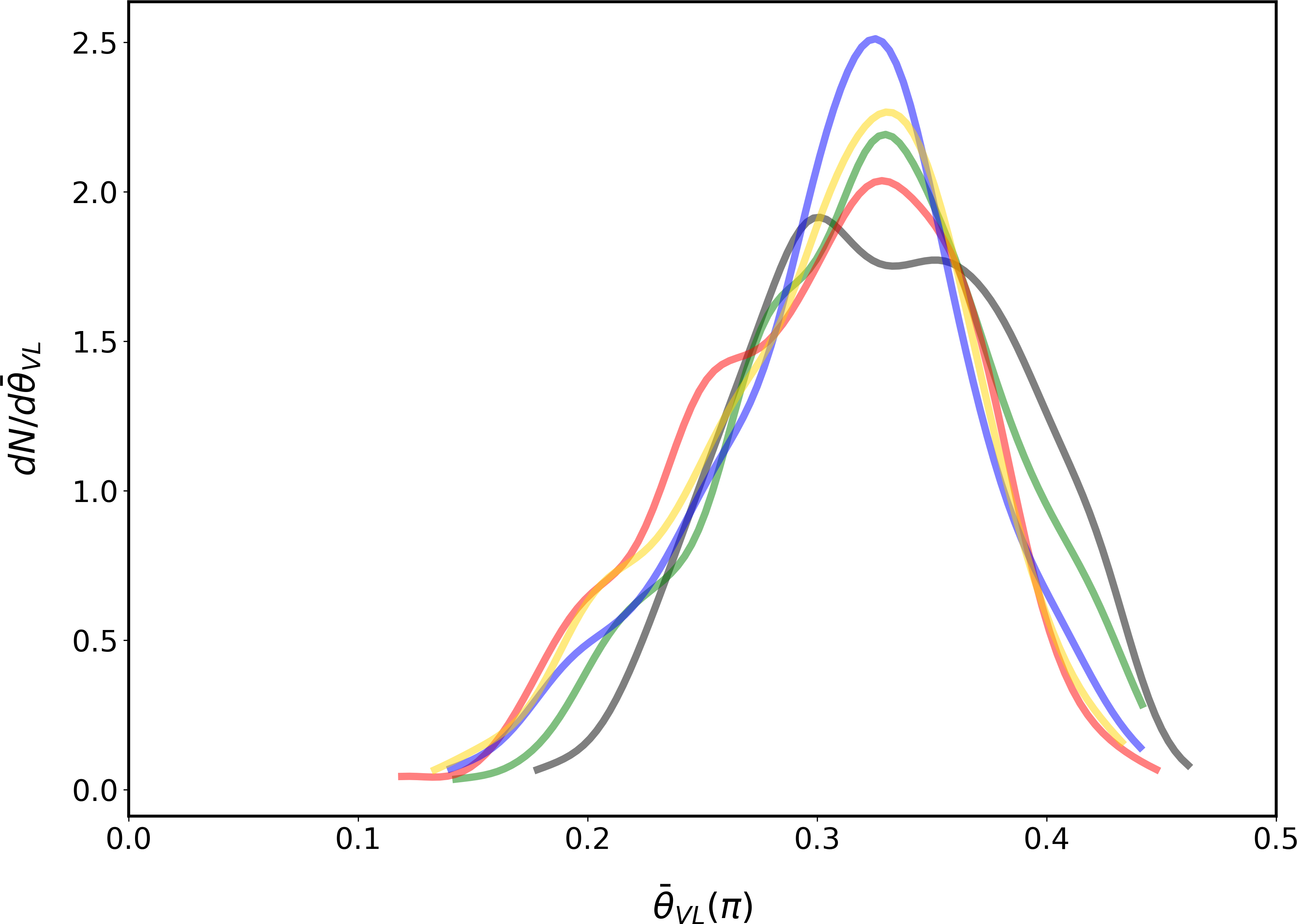}
    \caption{Density distribution functions of the mean angle $\overline{\theta}_{VL}$ between the largest principal axis and the internal velocity $v_{rms}$. The colour code is the same as in Fig. \ref{fig:thermaleq}}
    \label{fig:KDE_TH_VLoc}
\end{figure}

In Fig. \ref{fig:KDE_TH_V} we show the distributions of $\overline{\theta}_V$. This distributions are wider than those of Fig. \ref{fig:KDE_TH_B} and their modes do not show any apparent trend, however we can see that angles larger than $\pi /4$ are strongly favored, indicating that the internal velocity of the clumps is also closer to a direction perpendicular to the elongation of the clumps. For this angle we also show the resulting distribution for the B00S case, which has the strongest preference for $\overline{\theta}_V \sim \pi/2$. When the probability of having $ \pi/4 \leq \overline{\theta}_V  <   \pi/2$, $\mathcal{P}_{V}$ (second column in Table \ref{tab:angles}) is computed, this behavior is more visible. Although, the values of $\mathcal{P}_{V}$ are smaller than those of $\mathcal{P}_{B}$ and no clear trend with $B_0$ can be observed.  When the mean velocity of the clump is subtracted, i.e. when the bulk motion is excluded from the calculation (Fig. \ref{fig:KDE_TH_VLoc}), the preference  for large angles (>$\pi/4$) is lost. In this case all the density distributions peak slightly above or around $\pi/3$ and there is not a trend with $B_0$.  

\begin{figure}
    \includegraphics[width=\columnwidth]{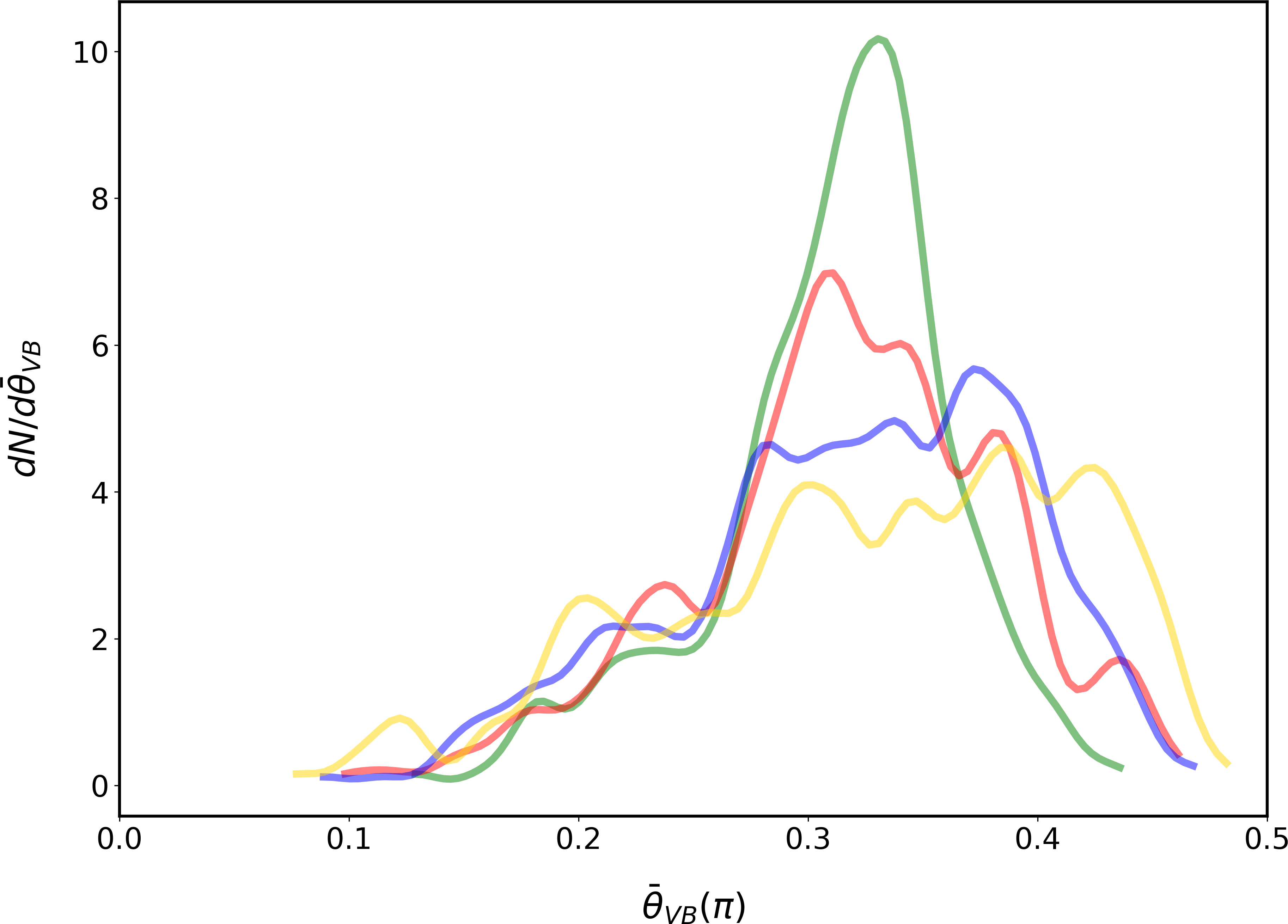}
    \caption{Density distribution functions of the mean angle $\overline{\theta}_{VB}$ between the velocity and the magnetic field of the clumps. The colour code is the same as in Fig. \ref{fig:crutcherplot}}
    \label{fig:KDE_TH_VB}
\end{figure}

Both $\overline{\theta}_B$ and $\overline{\theta}_V$ seem to have similar relative orientations relative to the clump's main axis, however, as those angles are not constrained to the same plane they can point in different directions in a three-dimensional system. This is clear when looking at the distributions of $\overline{\theta}_{VB}$, displayed in  Fig. \ref{fig:KDE_TH_VB}. The absence of clumps with low $\overline{\theta}_{VB}$ values shows that none of the clumps have velocities and magnetic fields pointing in the same direction. From the same figure, it can be seen that the B01S, B05S and B10S distributions have maximums at around $\pi/3$, while B20S has its maximum closer to $\pi/2$. Two trends are evident in this case: 1) as the initial magnetic field increases the small angle tail of the PDF reaches smaller values, indicating the presence of motions in a direction closer to that of $\mathbf{B}$; 2)  the distribution becomes wider as $B_0$ increases. In an attempt to quantify the behavior of this angle,  we compute probabilities for two intervals (fifth and sixth columns in Table \ref{tab:angles}). While the probability of having  $ \pi/4 \leq \overline{\theta}_{VB} <   3\pi/8$, $\mathcal{P}_{VBL}$, decreases drastically with $B_0$, the probability of motions with directions farther from that of the magnetic field, $\mathcal{P}_{VBH}$, increases substantially with $B_0$. Please note that $\overline{\theta}_{VB}$, unlike the two angles previously mentioned, involve the local variations of the two vectors. 
%
\section{Discussion}\label{sec:disc}
In this paper we worked on extracting and analyzing properties of clump like structures formed in forced (M)HD simulations that model the CNM of the local ISM with the intention of characterizing and comparing them between sets of clumps formed with different initial magnetic field intensities $B_0$. After we selected our clump samples we performed different analysis in order to obtain their physical, morphological, and geometrical characteristics. It is worth mentioning that during the data analysis reported in this work, we searched for correlations between the general physical properties of the clumps in our sample and its morphological characteristics and relative orientations but we have not found any significant correlation.  Even though only  a small number of the quantities that we analyse can be directly compared with observational results, see below, the aim of the work was to asses the expected physical properties of diffuse clumps forming in magnetized environments. This is specially relevant due to the lack of these kind of studies.
\subsection{Internal motions and magnetic field intensity}\label{secs:disc_machB}
Observationally, gathering information about CNM clumps is not trivial. In this kind of gas, the spin temperature measured from  the 21cm line can be used to trace the kinetic temperature  \citep{Heiles2003}, which in turn allows for estimations of the Mach number and the number density.  On the other hand, the Zeeman effect can be used to measure the magnetic field integrated along the line of sight. With this in mind, the position of a clump in the density vs. magnetic field ($n$ vs.$B$) diagram and its Mach number, obtained from our analysis, can be directly compared with observations.

The area occupied by our clump sample in the $n$ vs.$B$ diagram (Fig. \ref{fig:crutcherplot}) is consistent with the positions in this diagram of the clumps reported by \cite{HeilesTroland2004} for the solar neighbourhood. This is somehow natural considering our choice of the cooling function and the initial magnetic field strengths, both adapted to the local HI gas, and it gives us confidence that our clump sample reproduces, at least to some extent, the physical properties of the CNM clumps in the solar neighbourhood.  It is worth noting, however, that the clumps in our sample lie over the thermal equilibrium curve at pressures up to a factor of  3 above the mean neutral gas pressure at the solar galactocentric radius ($\overline{P}/k\sim 3000$~K cm$^{-3}$, \cite{Wolfire2003}), and up to a factor of 2 above the maximal pressure allowed for having two phases in pressure equilibrium, which for our cooling function is about 4900~K cm$^{-3}$ (Fig. \ref{fig:thermaleq}).  Although having clumps in thermal equilibrium is not surprising either, it is important to stress that for densities above $\sim 60$~cm$^{-3}$ ($T\lesssim 80$~K) the assumption of thermal pressure equilibrium at $\overline{P}$ is not appropriated.  As the estimation of the number density from the spin temperature requires an assumption about the thermal pressure, thermal equilibrium would lead to better estimations than pressure equilibrium.

As described in the previous section the absolute Mach number distribution peaks close to 4, in good agreement with observations \citep[see][and references therein]{Nguyenetal2019}, while the Mach numbers calculated with the $v_{rms}$ have peak values closer to one. This fact indicates that the clumps as a whole move supersonically through the medium while internally their movements are barely transonic. This is in accordance with previous numerical works \citep[e.g.][]{Heitsch_2005,HenAuMiv2007, Saury2014} and favors the interpretation of observed values as consequence of relative motions of different clumps along the line of sight. The fact that in our clump sample the contributions of bulk motions drastically differ from those of the internal motions may have implications on how the clumps get elongated, this will be discussed together with the geometrical properties of the clumps in section \ref{secs:disc_angles}. 
\subsection{Pressure balance}\label{secs:disc_press}
The plasma beta probability distribution of the clumps reveals that not all of the clumps in our population are magnetically dominated, in fact it takes the highest of our tested $B_0$ in order to get a probability distribution entirely in the $\beta_P < 1$ region (Fig. \ref{fig:betas}). For the low $B_0$ models, the $\beta_P$ distributions reach values above 10, indicating the presence of clumps where the magnetic pressure is highly subdominant.  Having a considerable fraction of clumps dominated by the thermal pressure indicates that most of our population should not be expected to follow the trends dictated by the initial magnetic tension, specially when considering the lower values of $B_0$. This may be ratified by the probability distributions of the Alfvén-Mach numbers which, similar to what happens with the sonic Mach number distributions, is different depending on whether the internal velocity $v_{rms}$ or the mean velocity including bulk motions is considered (Fig. \ref{fig:MachAs}). In the bulk distributions almost all of the clumps are clearly super-Alfvénic. In contrast, for the case in which only the internal velocity of the clumps is included most of the distribution modes lie in the sub-Alfvénic or trans-Alfvénic regime, excepting the one for the distribution coming from the low $B_0$ model. However, in this case all of the distributions have a super-Alfvénic region, which means that there is a group of clumps that internally have an important contribution on their dynamics due to internal motions. Using median values from observed velocity, temperature, and magnetic field intensity from the Millenium  Arecibo 21 Centimeter Absorption-Line Survey and assuming a typical CNM pressure $P/k=3000$~K cm$^{-3}$, \cite{HeilesTroland2005} estimate an Alfvénic Mach number of 1.3 for the local CNM. This value is more consistent with the distributions that we obtain when considering only internal motions, making it difficult to interpret the differences in the Alfvénic Mach number distributions in an analogous way as for the sonic Mach number. Unfortunately, and unlike the case of the sonic Mach number, there are not other observational estimates.
\subsection{Effects of the magnetic field on morphology}\label{secs:disc_shape}
Regarding the shape of the clumps, from the combined information of asphericity ($A_3$, Fig. \ref{fig:asphericity3d}) and prolatness ($S_A$, Fig. \ref{fig:prolatness}) it is certain that most of the clumps that we found in our models are filament-like structures. The only simulation that produced an important amount of non-filamentary clumps is the one without magnetic field. The probability of having more aspherical ($A_3\geq 0.75$) clumps increases for models with $B_0\neq 0$,  for which the probability of having slightly aspherical ($A_3\leq 0.25$) clumps is smaller than for the $B_0= 0$ model and decreases with $B_0$. Furthermore, the probability of forming highly prolate clumps ($1.25<S_A\leq 2$) is significantly larger for magnetized models while they present a lower probability of having oblate structures ($S_A<0$). The fact that even with a $B_0$ as small as $B \sim 0.4$\,$\mu$G the shape of the formed clumps is so qualitatively different than that of non-magnetized clumps implies that magnetic fields are heavily relevant to the structure of the neutral clumps of the ISM. These results are in accordance with previous works presenting magnetic mechanisms leading to the elongation of density structures in media where self-gravity does not dominate. In particular, \cite{Hennebelle2013} explained the origin of filaments in the HI gas as well as in the diffuse molecular gas in terms of the combined effects of the strain produced by turbulence and the confining effects of the Lorentz force; while more recently \cite{Xu_2019} argued that turbulence anisotropy  (i.e. turbulent mixing perpendicular to the ambient magnetic field) of Alfv\'enic turbulence is the source of filament formation in low density environments.  

However, it has also been suggested that apparent HI filaments are edge-on shells or sheets originated by shock-waves resulting from supernova explosions \citep{kalberlaetal2016ApJ...821..117K, kalberlalerphaudhav2017}. From the quantities that we use to evaluate the shape ($A_3$ and $S_A$ ), we are convinced that an aspherical non-filamentary clump is not necessarily a sheet-like structure as the aspect ratio of the clumps is not so different between samples, meaning that even the flatter structures are considerably broad along their minor axis. Note that the  possible presence of CNM sheets originated by large scale shocks, as the ones suggested by \cite{Heiles2003} and \cite{kalberlakerp16}, can not be evaluated from our models because its formation needs a spatially localized energy injection. Even if in our models the energy injection scale associated with the Fourier forcing, 50~pc, is comparable with the one reported in the literature \citep[see e.g.][]{Gent2013} for the neutral atomic ISM, the non-localized nature of this kind of forcing does not favour the formation of shell-like structures. It is thus likely that the absence of quasi-two-dimensional structures could be reverted with a more realistic forcing.
\subsection{Magnetic field alignments}\label{secs:disc_angles}
\begin{figure}
	\includegraphics[width=\columnwidth]{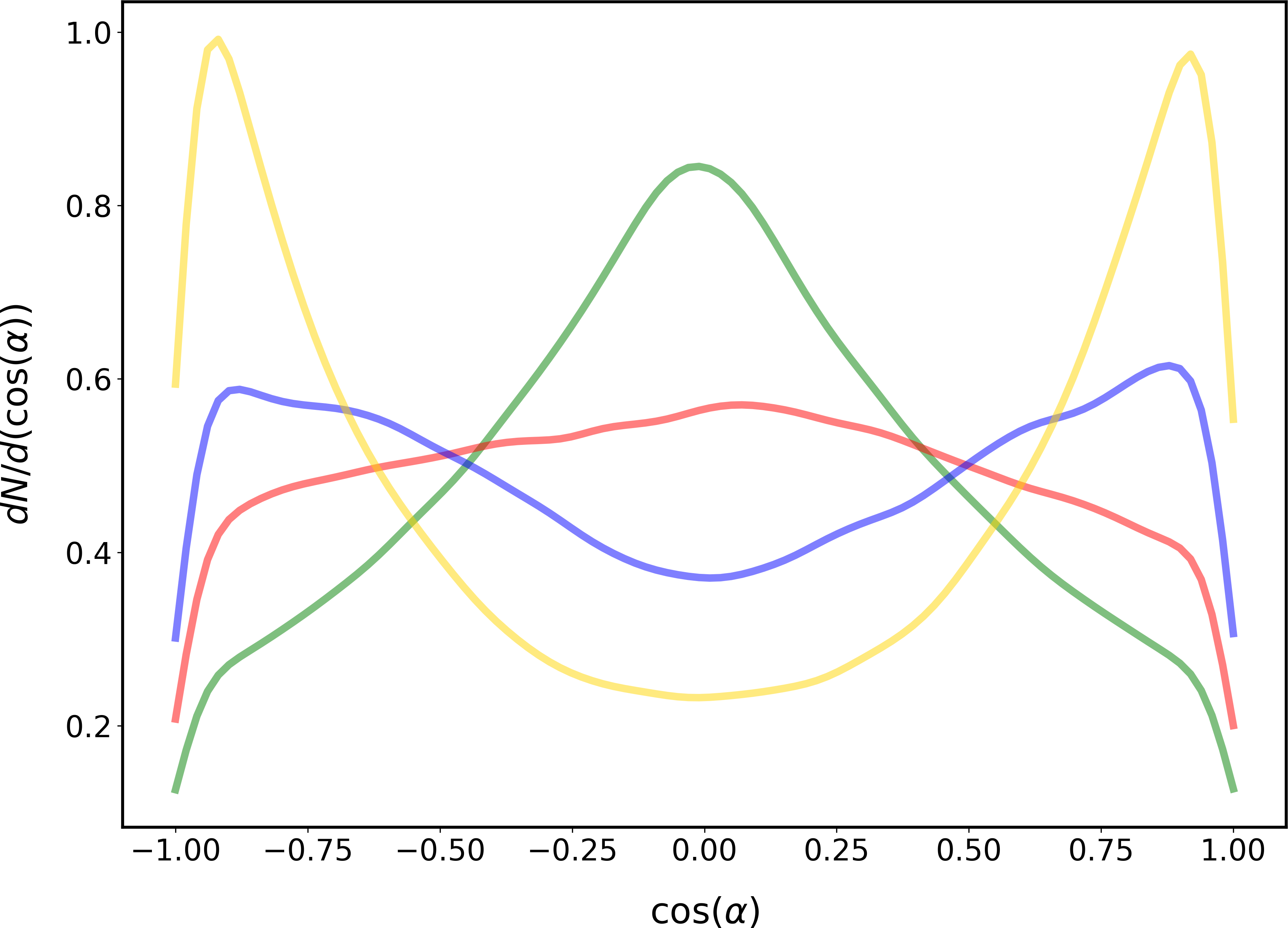}
    \caption{Mean density distribution functions of $\cos\alpha$, where $\alpha$ is the angle between the density gradient local magnetic field. The line colour code is the same as in Fig. \ref{fig:crutcherplot}.}
    \label{fig:KDE_ro}
\end{figure}

The calculated mean angle $\overline{\theta}_B$ between the local magnetic field and the largest principal axis shows that they are preferentially perpendicular to each other and that the preference is stronger for the strongest $B_0$ (Fig. \ref{fig:KDE_TH_B}). It is also worth nothing that the probability of finding a clump with an angle $\overline{\theta}_B = 0$ is practically zero. These results are partially  contrasting with our previous results, reported in \cite{Marco18} and found through the analysis of HROs, indicating that in the same  models the cold gas as a whole is predominantly aligned with the magnetic field and that the alignment is reduced as the initial magnetic field increases.
In order to understand the relationship between these two results, for all the clumps, we computed the PDF of $\cos\alpha$, where $\alpha$ is the angle between the density gradient $\nabla n$ and $\vec{B}$, i.e. a KDE version of the HRO\footnote{We also calculated KDE for each clump and a KDE of the modes of the distribution of each clump which resulted in similar results}. Let us remind the reader that when $\nabla n$ is preferentially perpendicular to $\vec{B}$,  i.e $\cos\alpha=0$, the density structure is preferentially parallel to $\vec{B}$.  In Fig. \ref{fig:KDE_ro}, we display the resulting distributions, where it can be seen that in clumps resulting from the low $B_0$ model the magnetic field is preferentially aligned with the density structures but this preference is gradually lost as $B_0$ becomes stronger. For the strongest $B_0$ the magnetic field is preferentially perpendicular to the density structures. The presence of this kind of structures, frequently associated with high column densities, has been often related with the effects of gravity \citep[e.g.][]{PlanckMol2016A&A...586A.138P}.  Our results are consistent with those obtained from shock compressed simulations by \cite{Inoue_2016}, who showed that they can also be formed in low density environments where gravity is not important.  

The differences between the magnetic field relative orientation obtained by the two different methods, imply that the relative alignment measured with the main principal axis captures the trend with $B_0$ but does not accurately describe the relative alignment between a density structure and its internal magnetic field. This could be due to curved filaments. The principal axes circumscribe the clumps in an ellipsoid whose largest elongation is in the direction of the largest principal axis. Even though the measures of the principal axes do help in determining how spherical and prolate or oblate are the structures they do not provide an insight on how curved a filament may be. For example, a filament may be contained by an ellipsoid whose largest principal axis is perpendicular to the velocity or to the magnetic field, but said filament may have curved limps that do not share the direction of the largest principal axis. Curved filaments are expected to be more abundant at low $B_0$ due to the lower magnetic tension. 

On the other hand, from Fig. \ref{fig:KDE_ro}, and in particular from the clear transition from parallel to perpendicular as $B_0$ increases, it is also clear that the results from examining all the CNM  as a whole \citep{Marco18} can not be translated unequivocally to individual clumps. This fact may be due, at least partially, to the relatively high density, compared to the densities of cold gas in the thermal pressure equilibrium regime, of selected clumps. 

Finally, from the same figure, it is worth noting that for individual CNM clumps in the same density range, the relative direction  of the internal magnetic field seems to be strongly determined by the ambient magnetic field. In \cite{Marco18}, we attributed the progressive loose of alignment as $B_0$ increases to a combined effects of the inhibition of the isobaric mode of the thermal instability in the direction normal to $\vec{B}$ and the enhancement of magnetic pressure. Note however, that the preferred  alignment of filaments from the low $B_0$ model suggests that the mechanism of alignment proposed by \cite{Inoue_2016} in the context of compressed layers, which is more effective as the shear strain in the direction of the magnetic field gets stronger, is presumably playing also an important role in these forced models.

In view of the limits of the principal axis approach to get the relative alignments, the interpretation of the distributions of the angle between the velocity and the largest principal axis is not straightforward. However, the difference on the distributions depending on which velocity, the  mean or the $rms$ velocity is used, i.e the fact that the angle between the mean velocity and the main axis tends to be nearest to $\pi/2$, suggests that compression facilitates the  accumulation of material in directions almost perpendicular to the flow.

From the behavior observed in the distributions of the angle $\theta_{VB}$ between local velocity and local magnetic field (Fig. \ref{fig:KDE_TH_VB}), which become wider and more skewed to high angles for the strongest $B_0$, we can assert that the internal motions in the clumps of our sample are not preferentially along the magnetic field lines.
\subsection{Two dimensional properties}\label{secs:disc_2d}

Our analysis of the relative alignment of clumps obtained from two-dimensional projections of our clumps (see Appendix \ref{sec:app}), rendered information about how the statistics associated to these  projections conserve information about the statistics from the three dimensional morphological and geometrical properties of clumps. 

In the two studied projections ($\parallel$ and $\perp$ to the initial magnetic field) the effect of magnetic field on producing more anisotropic structures is conserved.
In the perpendicular projection, a systematic increase with $B_0$ in the preference for highly elongated structures can even be observed. This suggests that at least part of the observed filament-like geometry in the HI gas could be due to the magnetic field. 

 When the line of sight of the clump is parallel to the initial magnetic field direction the probability distribution of the angle $\theta_B$ between the largest principal axis and the magnetic field of the clumps is dispersed through a wide angle range for all the $B_0\neq 0$ values. In this case, the mean component of the magnetic field lies outside the analyzed plane, hence a distribution of clumps aligned with multiple directions is expected. On the planes perpendicular to the initial magnetic field the probability distributions of $\overline{\theta}_B$ do actually present variations with $B_0$: they show the vectors as preferentially perpendicular in the B10S and B20S cases, and  not clear preferred relative direction for B05S, quite similarly to the three-dimensional results for these three models. This implies that for sufficiently magnetized environments, the two-dimensional objects, and its associated magnetic field, statistically conserve the alignment properties of three-dimensional objects. However, the important fluctuations in the distributions together with the fact that for LOS$\parallel \vec{B_0}$ no preferred angle is expected, suggest that for general orientations with respect to the mean magnetic field this correspondence could be lost. 

The two-dimensional analysis that we present is not intended as a synthetic observation. Exploring in a deep way how the misalignment between the density and the magnetic field that we find for strongly magnetized models is translated in mock observations, what observational consequences should we expect from this behavior, and how these are related to recent observational results showing an apparent preference for an alignment between the CNM density structures and the magnetic field \citep[e.g.][]{Clark2014ApJ...789...82C, PlanckHI2016A&A...586A.135P,kalberlaetal2017A&A...607A..15K,jelicetal2018},  deserve a more careful study which is out of the scope of this work.    

\section{Conclusions}\label{sec:conc}

In this paper we studied the cold clumps formed via thermal instability in MHD simulations of the atomic interstellar medium. Our main interest where to classify the clumps according to their geometry, to quantify their physical properties, and to search for alignments between them and the magnetic field. We worked with simulations that had initial magnetic fields with values: $0.4$, $2.1$, $4.2$ and $8.3$\,$\mu$G. Additionally we worked with the clumps formed in a simulation with a similar setup but without a magnetic field. We used the principal axes of the clumps to measure their aspect ratio, asphericity and prolatness. We built probability density distributions of the properties of the clumps via Kernel Density Estimations. Finally we used directional statistics to look for the mean angle between the largest principal axis of the clumps and the velocity and the magnetic field. We do not find any significant correlation between the general physical properties of the clumps and their morphological descriptors and relative alignments and neither between the latter two. Our results drove us to formulate the following conclusions: 

\begin{itemize}
    \item The morphology of clumps formed in hydrodynamical simulations is intrinsically different to to the morphology of those formed in MHD simulations. This occurs even when the initial magnetic field of the MHD simulations is weaker than the observationally estimated values for the local diffuse ISM. 
    
    \item The clumps on all of our tested models are predominantly filament-like structures, but we find a quantifiable enhancement of the asphericity  as well as a quantifiable tendency to form more highly prolate clumps for all the MHD models with respect to the purely hydrodynamic case. As a consequence, it seems that conclusions about the CNM structure based non-magnetic results should be taken with caution. Additionally, the simulations with the highest initial magnetic field are more prone to form more aspherical clumps. This tendency can be recovered in a two-dimensional projection perpendicular to the initial magnetic field.
    
    \item For a clump sample with physical properties comparable with those reported by observations, we find a wide variety of regimes concerning the pressure balance and the Alf\'enic Mach number. This could imply that describing CNM clouds by single values of $\beta_P$ and/or $M_A$ might constitute an over-simplification which could potentially exclude relevant physics.

    \item The mean magnetic field is more influential to the relative orientation of the analyzed clumps than to any other property. On one side, the elongation direction of clumps, as determined by its larger principal axis, has preferred angles between $\pi/4$ and $\pi/2$ with respect to the internal magnetic field. The preference for near to perpendicular relative orientations gets accentuated for the samples with a more intense ambient field. On the other side, the relative orientation between the local magnetic field and the local density structure, as measured by the density gradient, shifts from mainly parallel to mainly perpendicular as the  ambient field increases.
    
    \item We introduced the use of the asphericity and the prolatness to characterize the morphological properties of density structures. These attributes appear to be more sensitive to morphological variations than the traditional aspect ratios and allowed us to successfully quantify differences among samples resulting from different models. Estimating the two-dimensional versions of these quantities for observed clumps, both synthetic and natural, could bring about insight on the real CNM structures.   

\end{itemize}

\section*{Acknowledgements}
MV is a doctoral fellow of CONICET, Argentina. AG acknowledges partial support from the CONACYT grant 255295.  Both
authors appreciate the support of UNAM-DGAPA, PAPIIT through
the grant IN111318.  Both authors would like to thank an anonymous referee for his/her suggestions and comments, which helped to improve the quality of the manuscript. This work has made extensive use of the NASA’s Astrophysics Data System Abstract Service.

\section*{Data Availability}
The data underlying this article will be shared on reasonable request to the corresponding author.

\bibliographystyle{mnras}
\bibliography{references} 


\appendix
\section{Two-dimensional properties}\label{sec:app}
\begin{figure}
	\includegraphics[width=\columnwidth]{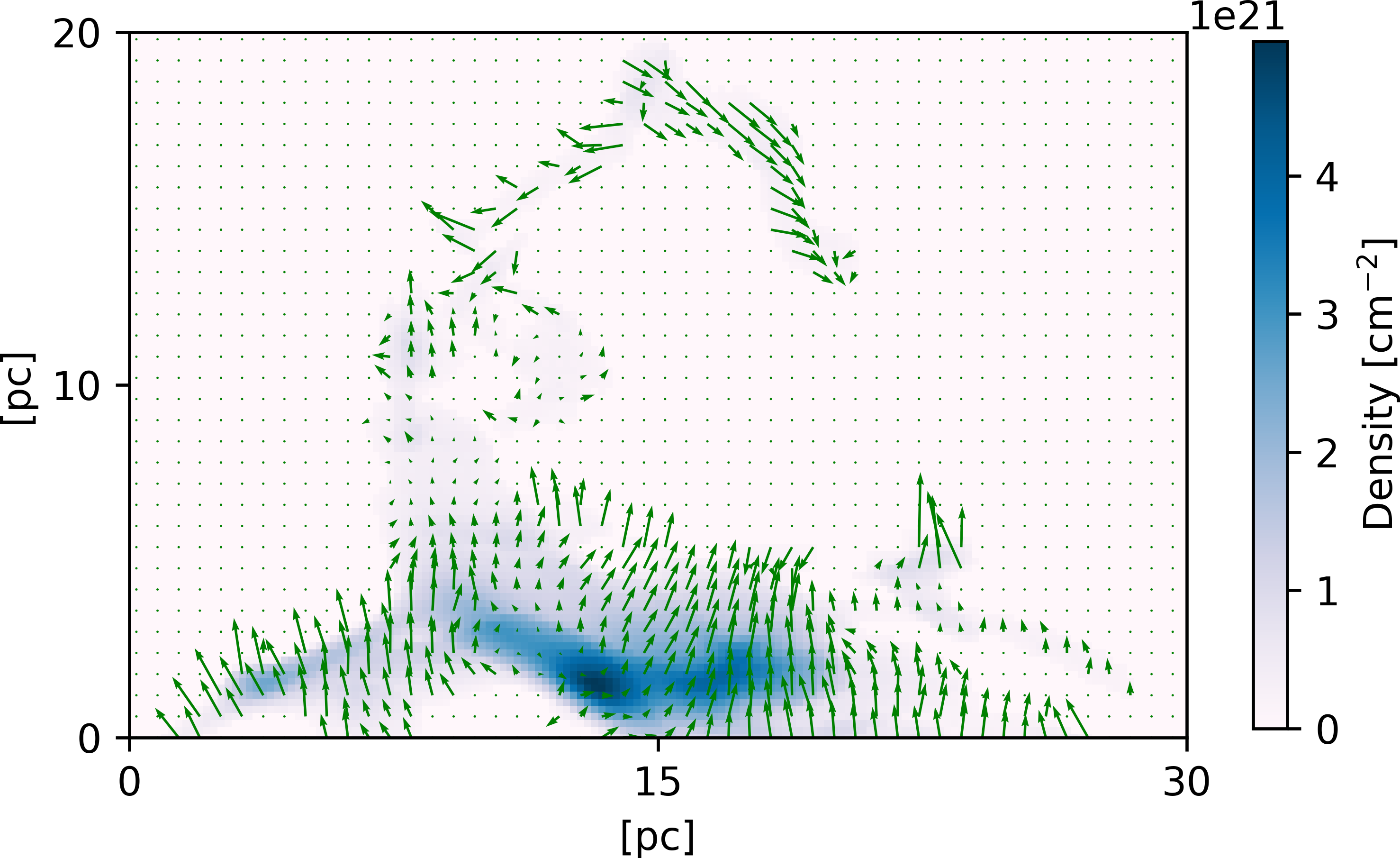}

	\includegraphics[width=\columnwidth]{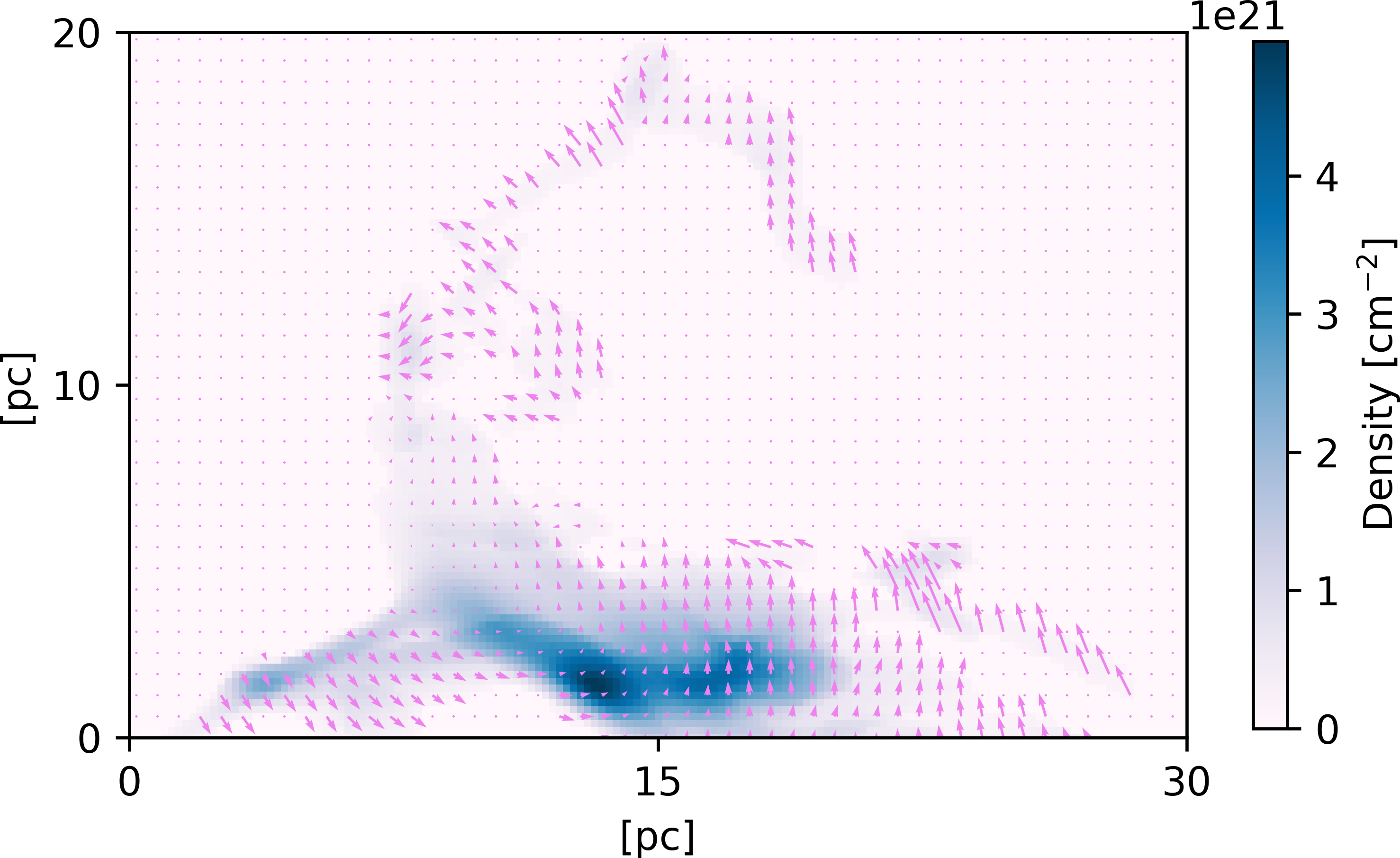}
    \caption{Sample two-dimensional clump selected with the lower density threshold. The superimposed vectors in each panel are the local  magnetic field (green, top) and  the local velocity field (purple, bottom).}
    \label{fig:2d_clump_B01}
\end{figure}

As the information available from observations is obtained from two-dimensional objects it is interesting to study weather or not the findings of our three dimensional clumps have an equivalence in two dimensions. In order to explore this possibility, we created projections of data cubes containing only the voxels selected in 3D clumps with $n>50$~cm$^{-3}$. In those projections we searched and selected clumps with surface densities above three thresholds: $1 \times 10^{20}$\,cm$^{-2}$, $5 \times 10^{20}$\,cm$^{-2}$ and $1 \times 10^{21}$\,cm$^{-2}$, which are consistent with observed values for the CNM gas \citep[see][]{Nguyenetal2019}. Additionally, we limit the clump area range by considering only clumps with areas within a minimum of 100 pixels (equivalent to a circle with $r \sim 2.5$\,pc) and a maximum of 15000 pixels (equivalent to a circle with $r \sim 30.5$\,pc). An example of a two-dimensional clump is shown in Fig. \ref{fig:2d_clump_B01}, where the projected magnetic field (top) and velocity (bottom) are superimposed on the surface density map. 
We use three different directions parallel to the cube axes, thus obtaining information at the planes YZ, XZ, and XY which result from the integration of the dense cubes along the $\mathbf{x}$, $\mathbf{y}$, and $\mathbf{z}$ axis, respectively. It serves as a brief reminder that the initial magnetic field of the simulations is along the $\mathbf{x}$ axis. As in the 3-dimensional case, the density distributions that we show combine results from the three thresholds mentioned above and for the  snapshots described in section \ref{sec:results}.
Due to the construction method of the two-dimensional maps, two dimensional objects selected from projections can not be directly identified with the three dimensional clumps analyzed in section \ref{sec:results} as they may be composed by parts of different three-dimensional clumps. This creates a difficulty when evaluating the results of searching for the 2-dimensional equivalent quantities:  asphericity eq.\,(\ref{eq:aspher_2d}), aspect ratio, and angles between vectors eq.\,(\ref{eq:theta}). 

\begin{figure}
	\includegraphics[width=\columnwidth]{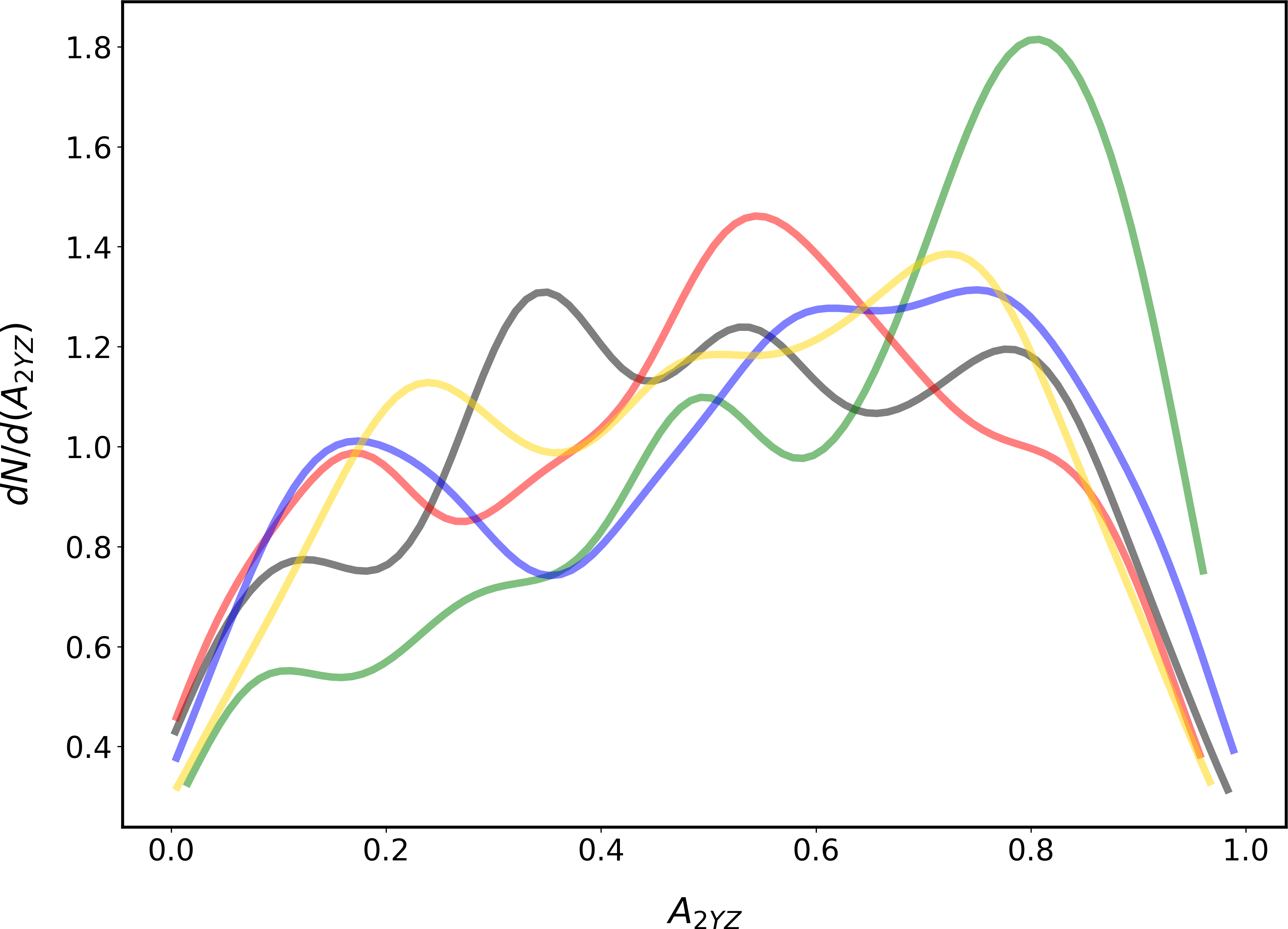}
    \caption{Density distribution functions of the two-dimensional asphericity, $A_2$, in the $YZ$  plane. The colour code is the same as in Fig. \ref{fig:crutcherplot}}
    \label{fig:A2_YZ}
\end{figure}

\begin{figure}
	\includegraphics[width=\columnwidth]{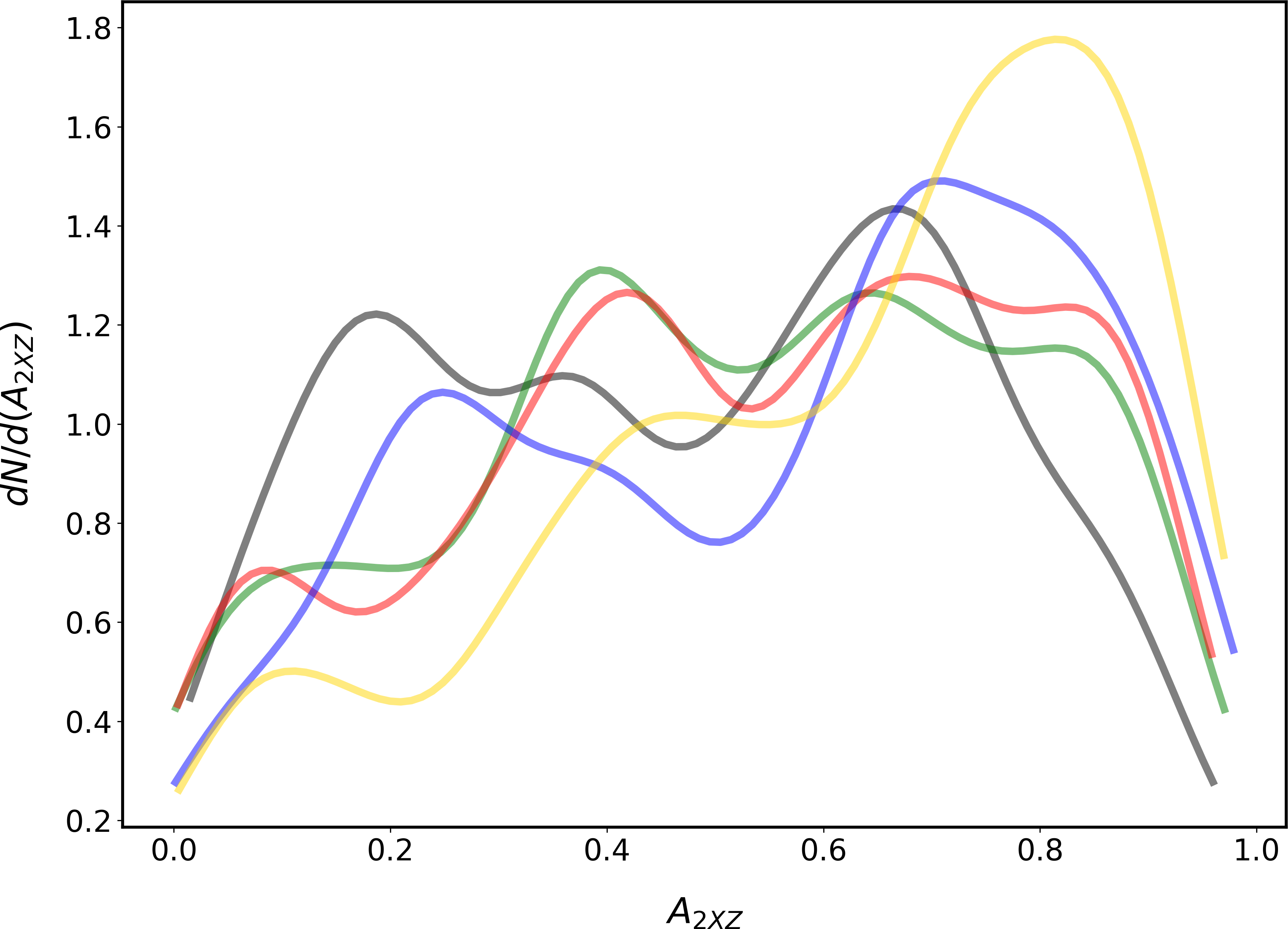}
    \caption{Density distribution functions  of the two-dimensional asphericity, $A_2$, in the $XZ$  plane. The colour code is the same as in Fig. \ref{fig:crutcherplot}}
    \label{fig:A2_XZ}
\end{figure}

Figures \ref{fig:A2_YZ}  and \ref{fig:A2_XZ} display the  density distribution function for the two dimensional asphericity, $A_2$ (equation (\ref{eq:aspher_2d})), resulting from two different projections; one along the initial magnetic field direction ($YZ$ plane) and another one along a direction perpendicular to this field ($XZ$ plane), respectively. For the projection along $\vec{B_0}$, the distribution resulting from the magnetized model with the lowest magnetic field intensity, B01S, shows a clear preference for more anisotropic structures, while  for the perpendicular projection  is the highly magnetized model, B20S, which has this behavior. In the latter case, it can also be noticed a clear decline of preference for structures with $A_2\gtrsim 0.8$ as $B_0$ increases.  These results show that projected clumps maintain, at least to some extent, the morphological effects of the magnetic field presence (see section \ref{sec:results}). %

\begin{figure}
	\includegraphics[width=\columnwidth]{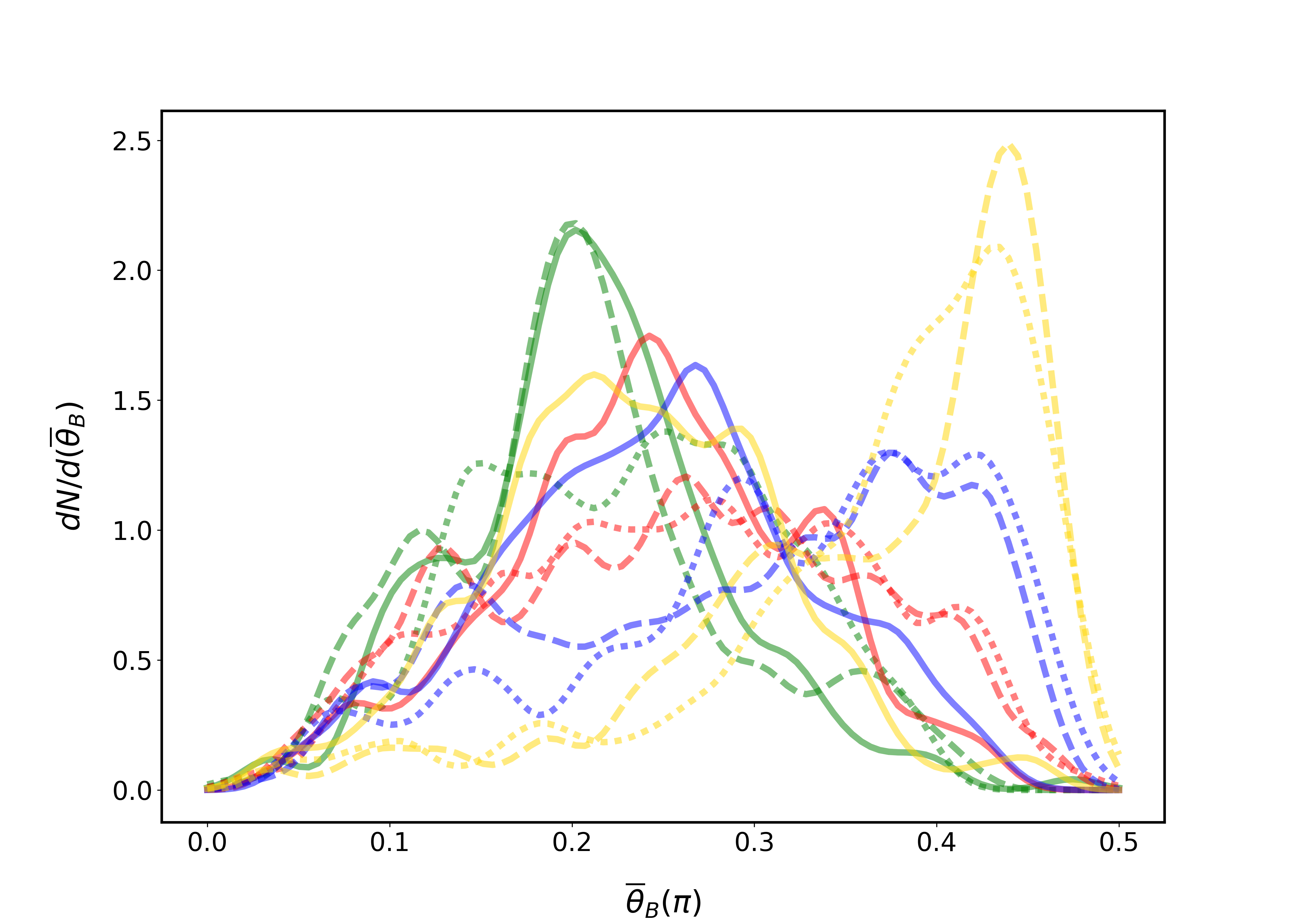}
    \caption{Density distribution functions of the 2D mean angles $\overline{\theta}_{B}$, between the largest principal axis and the internal magnetic field, in the $YZ$ (solid line), $XZ$ (dashed line), and $XY$ (dotted line) planes. The colour code is the same as in Fig. \ref{fig:crutcherplot}}
    \label{fig:TH_B_2D}
\end{figure}

In Fig. \ref{fig:TH_B_2D} we plotted the probability distribution of the angle $\overline{\theta}_B$ between the largest principal axis, which we use as a proxy of clump orientation, and the internal magnetic field in the three analyzed planes. From that plot it is clear that also in two dimensions, the initial magnetic field intensity has a noticeable effect on the relative orientation between the clump and the magnetic field. In the $YZ$ plane (solid line) a peak at $\overline{\theta}_B\sim 0.2\pi$ can be distinguished for the model B01. As $B_0$ increases the peak in the angle distribution shifts to larger values and becomes less pronounced. In fact,  for B20 there is no preferred value of  $\overline{\theta}_B$. In the other projected planes (dashed and dotted lines), as $B_0$ increases the distribution changes from having a preferred low value or not showing a preferred value to a clear preference for almost perpendicular orientations. This behaviour is consistent with the results from \cite{Marco18}, where HRO from projections perpendicular to the initial magnetic field show a trend towards a less aligned magnetic field as $B_0$ increases. Note however, that none of the
selection criteria for projections presented in  \cite{Marco18} match the gas selected in the present work.

\begin{figure}
	\includegraphics[width=\columnwidth]{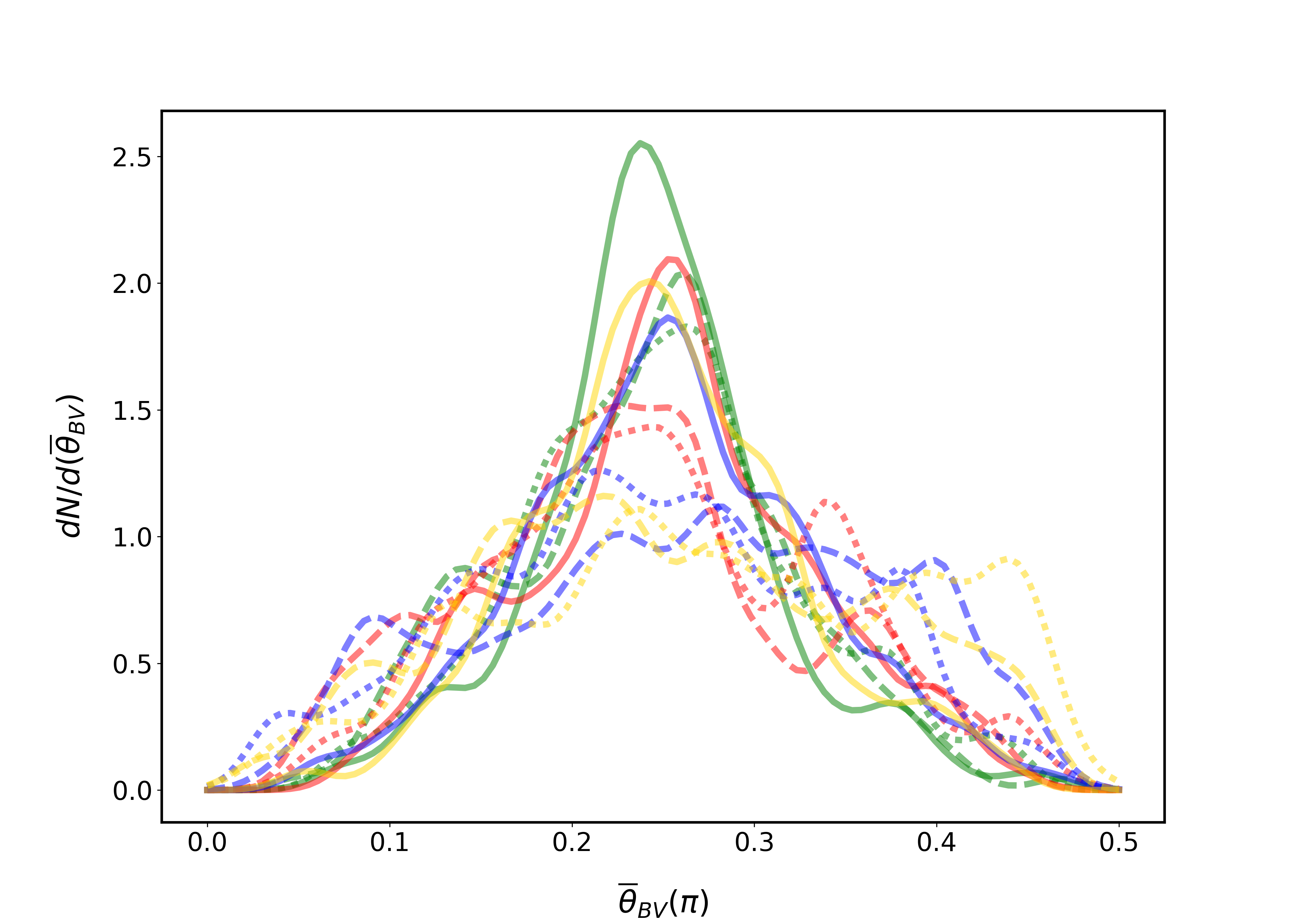}
    \caption{Density distribution functions of the 2D mean angles $\overline{\theta}_{BV}$, between the local magnetic field and the local velocity, in the $YZ$ (solid line), $XZ$ (dashed line), and $XY$ (dotted line) planes. The colour code is the same as in Fig. \ref{fig:crutcherplot}}
    \label{fig:TH_BV_2D}
\end{figure}

The probability distribution of the mean angle between the local magnetic field and the local velocity inside the 2D clumps, $\overline{\theta}_{BV}$, is displayed in Fig. \ref{fig:TH_BV_2D}. As expected, also for this angle the effects of changing the projection direction are larger for highly magnetized models. At the $YZ$ plane (solid line), the four magnetized models show a preference for  $\overline{\theta}_{BV}\sim \pi/4$.On the other planes, where the projection direction is perpendicular to the initial magnetic field,  this preference is maintained for the B01S model but is gradually lost as $B_0$ increases. In fact,  for the two models with the largest initial magnetic field, no preferred relative orientation can be distinguished.
\bsp	
\label{lastpage}
\end{document}